\documentclass[aps,pra,showpacs,twoside,10pt,floatfix,nofootinbib,twocolumn,longbibliography]{revtex4-2}
\usepackage[colorlinks=true, citecolor=red, urlcolor=blue ]{hyperref}
\usepackage{epsfig,newlfont,amssymb,amsfonts,amsmath,bm,subfigure,palatino,mathtools,amsthm,braket,soul,enumitem,color,comment,geometry,xcolor}
\usepackage[normalem]{ulem}
\definecolor{addcolor}{HTML}{7117C1}
\newcommand{\JK}[1]{{\color{black}#1}}

\begin{document}

\title{Logarithmic growth of peripheral entanglement concentrated via noisy measurements in a star network of spins}
\author{Jithin G. Krishnan, Harikrishnan K J, Amit Kumar Pal}
\affiliation{Department of Physics, Indian Institute of Technology Palakkad, Palakkad 678 623, India}
\date{\today}

\begin{abstract}
In a star-network of qubits interacting via Heisenberg interaction of  XYZ-type, we demonstrate a logarithmic growth of the localizable bipartite peripheral entanglement with increasing periphery-size and vanishing xy-anisotropy. This feature disappears when xy-anisotropy becomes non-zero, exhibiting an anisotropy effect, which can be negated by taking the system out of equilibrium by a qubit-local  magnetic field. In the large-center  and the competing-center limits of the model, the behaviour of LBPE is qualitatively different from that of the large-periphery limit. Also, the bipartite peripheral entanglement computed via a partial trace-based approach behaves qualitatively similarly to the LBPE in the large periphery limit, while in the other two limits, it behaves differently. We further consider the generalized description of localizable entanglement using unsharp measurements, and demonstrate that the logarithmic growth of LBPE is present for all noise strengths in the large-periphery limit, while in the competing-center limit, it does not.
\end{abstract}

\maketitle

\onecolumngrid

\section{Introduction}
\label{sec:intro}

Entangled quantum states~\cite{horodecki2009} have been shown to be a key resource for several quantum technological applications including quantum communication~\cite{bennett1992}, quantum cryptography~\cite{ekert1991}, quantum simulation~\cite{Dalmonte2018}, quantum metrology~\cite{Riedel2010}, measurement-based quantum computation~\cite{raussendorf2001}, and quantum algorithms~\cite{Bruss2011}. Being sources of entangled states, quantum many-body systems~\cite{amico2008} have been identified as candidate systems for implementing quantum information theoretic applications. Among the variety of quantum many-body systems, interacting quantum spin models~\cite{amico2008} have attracted bulk of the attention due to their natural representation of multi-qubit or multi-qudit systems pertinent to quantum protocols. The last two decades have also witnessed the implementation of a variety of quantum spin Hamiltonians in the laboratory using trapped ions~\cite{leibfried2003}, nuclear magnetic resonance systems~\cite{Vandersypen2005}, ultra-cold atoms and optical lattices~\cite{Duan2003}, and solid-state systems~\cite{Schechter2008}, thereby making the certification of the theoretically predicted entanglement properties of these systems possible. From this motivation, interface of quantum information theory and quantum many-body systems has emerged as a vibrant field of research advancing along two complementary directions.  On one hand, quantum technological applications have been implemented using interacting quantum spin systems, eg. quantum state transfer through spin chains~\cite{Bose2003} and one-way quantum computation~\cite{raussendorf2001} using cluster states~\cite{hein2006}.  On the other hand, quantum information-theoretic concepts have been extensively used to probe quantum many-body systems~\cite{amico2008}, leading to the introduction of projected entangled pair states~\cite{Schollwock2005}, and multiscale entanglement renormalization ansatz~\cite{Verstraete2004b}.

\begin{figure*}
\centering
\includegraphics[width=0.9\textwidth]{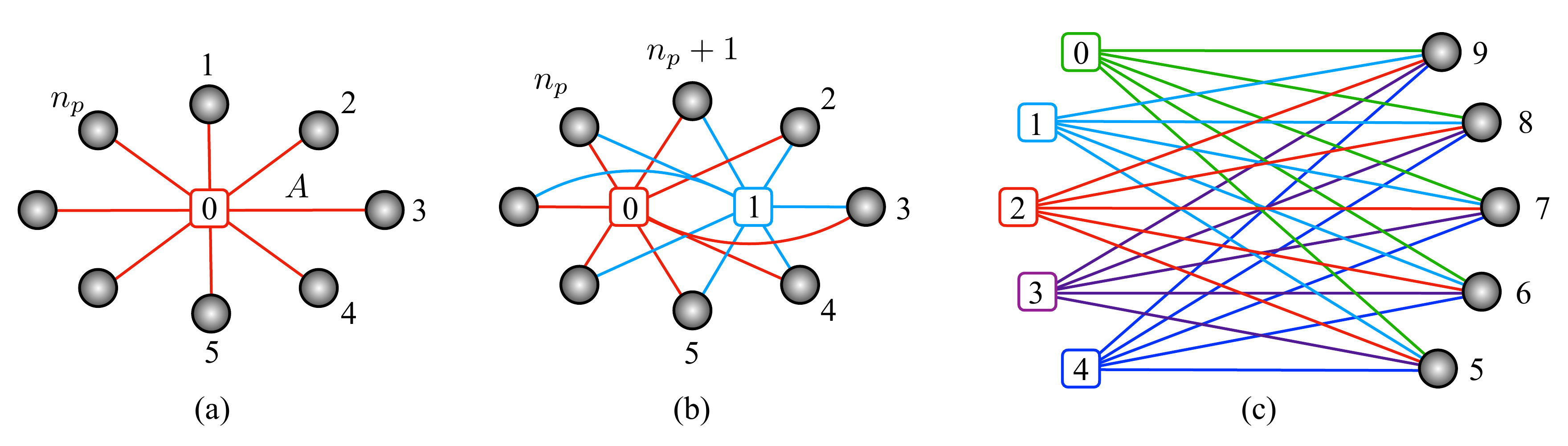}
\caption{(Color online.) Star networks of $n=n_0+n_p$ spin-$\frac{1}{2}$ particles with (a) $n_0=1$ and (b) $n_0=2$, where peripheral spins are represented by shaded circles. As $n_p$ increases, the large periphery limit $n_0/n_p\ll 1$ is achieved, while an interchange of the central spins with the peripheral spins (i.e., an interchange between $n_0$ and $n_p$) provides the limit $n_0/n_p\gg 1$. The equal periphery limit of the system with $n_0/n_p=1$ is demonstrated with $n_0=n_p=5$ in (c). In these star networks, each spin in the periphery interacts with each of the central spin with XYZ interaction.}
\label{fig:system}
\end{figure*}

Although quantum spin Hamiltonians defined on one- and two-dimensional regular lattices have primarily been in focus~\cite{amico2008}, achievement of control over the connection between any two qubits in an array realized in different substrates~\cite{Kane1998} have also given rise to  studies on quantum spin systems having a lattice structure modelled for specific purposes.   Prime examples along this line of studies are (a) a star network of spins~\cite{Hutton2004}, where a number of  \emph{central spins} are surrounded by a number of  \emph{peripheral spins} such that each of the central spin interacts with all of the peripheral spins (see Fig.~\ref{fig:system} for examples with one and two central spin(s)), (b) a star-network of spins where the peripheral spins also interact among each other~\cite{Richter_1994}, and (c) a star-chain network of spins, where a number of spin-chains have a common boundary spin~\cite{Yao2011}. While such a network was originally envisioned for achieving control over entanglement through one, or a group of \emph{preferred} (central) spins~\cite{Hutton2004}, they have also been used as switch in quantum networks~\cite{Yung_2011}, in quantum state transfer and cloning~\cite{Deng_2008}, in magnetic resonance imaging~\cite{Sushkov2014}, in studying measurement uncertainty~\cite{Haddadi2021}, and to implement quantum heat engines~\cite{Turkpence_2017}, refrigerators~\cite{Arisoy2021}, and quantum batteries~\cite{Liu2021} \JK{and recently in the studies on Quantum Darwinism \cite{Deffner2024}}.  Furthermore, the design of variants of a spin-star structure using semiconductor quantum dot~\cite{Zhang2006}, nitrogen-vacancy center in diamond~\cite{Zhao2011}, and in a cluster of seven dipolar-coupled nuclear spins of benzene molecules in a liquid-crystalline matrix~\cite{Lee2005}, along with the possibility of realizing theoretical spin-Hamiltonians in the laboratory using trapped ions~\cite{leibfried2003} and ultra-cold atoms and optical lattices~\cite{Duan2003} have made the experimental verification of the theoretical results on quantum spin-star systems a possibility.

Entanglement properties in a star network of spins have been extensively explored~\cite{Hutton2004,Anza_2010,Karlova2023,Stockton2003,Haddadi2019}, along with the quantum correlations not belonging to the entanglement-separability paradigm~\cite{Radhakrishnan2019,Haddadi2019} (cf.~\cite{modi2012}). Apart from the genuine multiparty entanglement over the spin-star system~\cite{Anza_2010,Karlova2023}, attention is also given to the entanglement over a spin-pair shared by the center and the periphery as well as a spin pair in the periphery~\cite{Hutton2004, Karlova2023}, and the entanglement over all possible bipartite split of the system~\cite{Stockton2003}. It has also been shown that the bipartite entanglement over an equal bipartition of the peripheral spins exhibits a logarithmic growth with the periphery-size~\cite{Stockton2003}.  Tracing over the central qubit(s) in the spin-star system leads to  a density matrix on the peripheral spins that can be written using the permutationally symmetric Dicke basis~\cite{dicke1954}, offering advantages in calculation. This has motivated Refs.~\cite{Hutton2004,Anza_2010,Karlova2023,Stockton2003,Haddadi2019} to rely mainly on a partial trace-based approach to compute entanglement over a chosen subsystem of peripheral qubits in the spin-star system~\cite{Hutton2004,Anza_2010,Karlova2023,Stockton2003,Haddadi2019}.  

An alternative approach for quantifying entanglement in a subsystem of a multi-qubit system is participation from the rest of the system via measurements, such that a maximum average entanglement over the subsystem in focus can be computed, referred to as the \emph{entanglement of assistance}~\cite{divincenzo1998}. Using this route, and performing independent local projection measurements on all qubits outside the chosen subsystem, \emph{localizable entanglement} (LE)~\cite{verstraete2004,sadhukhan2017,Mondal2024} over the qubits in the subsystem is computed. Localizable entanglement has been used for investigating the correlation length in one-dimensional quantum spin models~\cite{verstraete2004}, in quantifying entanglement over subsystems of stabilizer states~\cite{hein2006,amaro2018}, for detecting topological-to-non-topological quantum phase transitions~\cite{HK2022} in topological quantum codes, to characterize phases in 1D quantum spin models~\cite{skrovseth2009}, and in percolating entanglement through quantum networks~\cite{acin2007}. In the spin-star system, it is shown that maximally entangled states over a chosen qubit-pair in the periphery can be created by performing $\sigma^z$ measurements on all other qubits~\cite{Hutton2004}. However, the localization of entanglement over an arbitrary bipartition of the peripheral spins in a star network of spins via measurements on the central spins, which is in alignment with the original motivation of studying the star network of spins,  remains unexplored, and is investigated in this paper. Specifically, we ask the three following questions: 
\begin{enumerate}
    \item[(a)] What would be the static and dynamical behaviour of entanglement localized over equal bipartitions of the peripheral spins via \emph{sharp} (noiseless) projection measurements on the central qubit(s) against the size of the system?  
    \item[(b)] Do the features of entanglement computed via the measurement-based approach differ from those corresponding to the entanglement computed using the partial trace-based approach?
    \item[(c)] If the measurements on individual qubits are \emph{unsharp} (noisy), how would the results obtained in the cases of sharp projection measurements would change?
\end{enumerate}


Towards answering these questions, we consider a star-network of $n=n_0+n_p$ spins with $n_0$ central and $n_p$ peripheral spins, where each central (peripheral) spin interacts with each peripheral (central) spin via a fully anisotropic XYZ interaction~\cite{korepin_bogoliubov_izergin_1993}. We consider three limits of the model --  (1) the \emph{large periphery limit} given by $n_0/n_p\ll 1$, (2) the \emph{large center limit} given by $n_0/n_p\gg 1$,  and (3) the \emph{competing center limit}, represented by $n_0/n_p\rightarrow 1$. In the large periphery limit, for non-zero values of the $xy$- and the $z$-anisotropy parameters, we investigate the structure of the ground state of the system. We show that for odd $n$, the ground states are doubly degenerate for all values of the anisotropy parameters.  However, for even $n$, there is a finite energy gap between the ground and the first excited states for low and moderately high $n$. This energy gap decreases as $\exp{(-b n_p^m)}$ for $\gamma=0$, and as $bn_p^{-m}$ for $\gamma\neq 0$ with increasing $n_p$, and the ground state becomes \emph{effectively} doubly degenerate with a negligible energy gap once a \emph{critical} size $n_p^c$ of the periphery (and hence, a critical size $n_c$ of the system) is achieved. We also discuss the subtle changes in the ground state degeneracy in the $n_0/n_p\gg 1$ and $n_0/n_p\rightarrow 1$ limits of the model, as compared to the $n_0/n_p\ll 1$ limit.

We first study the interplay of the \emph{localizable bipartite peripheral entanglement} (LBPE) and the size of the partitions in the zero-temperature and in all three size-limits of the system.  In the case of the large center limit, we show that the LBPE over the equal bipartition of the peripheral spins exhibits a growth as $\log_2 n_p$ with the periphery-size when the system has no $xy$-anisotropy (cf.~\cite{Latorre2005} for a similar finding in  the Lipkin-Meshkov-Glick (LMG) model), while for non-zero $xy$-anisotropy, this feature is absent. We refer to this as the \emph{anisotropy effect}. Moreover, for a fixed system-size $n$ with a partition-size $n^\prime$, LBPE varies as $a\left[1-\exp\left(-b(n^\prime)^m\right)\right]$.   We also consider taking the  system out of equilibrium by turning on a magnetic field of constant strength on all spins. As the system evolves, we calculate the LBPE as a function of time at different points in the parameter space of the $xy$- and $z$-anisotropy parameters. The time-averaged LBPE in the long-time limit are shown to have a logarithmic growth with $n_p$ for all values of the $xy$-anisotropy parameter, implying a negation of the anisotropy effect due to the field-induced dynamics. Moreover, similar to the static scenario, average LBPE varies as $a\left[1-\exp\left(-b(n^\prime)^m\right)\right]$ for a fixed $n_p$. We also compute the bipartite entanglement over the equal bipartition of the peripheral spins using the partial trace-based approach, where the central spins are traced out of the state of the system, leading to a mixed state on the peripheral spins which is used to compute the bipartite entanglement~\cite{horodecki2009}. In the case of the large periphery ($n_0/n_p\ll 1$) limit, our analysis suggests a qualitatively similar behaviour of the partial trace-based entanglement with the system size, as that of the LBPE. \JK{Further, for moderate-sized systems, we demonstrate that the results regarding the bipartite peripheral entanglement computed using partial trace are robust against the presence of disorder in the spin-spin interaction strengths.}

Next, we consider the $n_0/n_p\gg1$ and $n_0/n_p\rightarrow 1$ limits of the spin-star system, and find that while the behaviours of LBPE against $n_p$ in the limit $n_0/n_p\rightarrow 1$ is qualitatively similar to that in the limit $n_0/n_p\ll 1$, in the limit $n_0/n_p\gg 1$, the trend is different. In the latter limit, the LBPE decreases monotonically and approaches zero asymptotically with increasing $n_0$ for $\gamma\neq 0$, while for $\gamma=0$, LBPE exhibits a saturation as $n_0$ increases, thereby exhibiting a different behaviour with varying system size than the results obtained for $n_0/n_p\ll 1$. Further, in stark contrast with the case of $n_0/n_p\ll 1$, in the large and competing center limits, the partial trace-based bipartite peripheral entanglement exhibits a trend different from that of the LBPE against the system size. In the large center limit,  the partial trace-based entanglement decreases monotonically with increasing $n_0$ as $\sim bn_0^{-m}$, and vanishes asymptotically, while $n_p$ is kept fixed. On the other hand, for $n_0/n_p\rightarrow 1$, the partial trace-based bipartite peripheral entanglement saturates with $n_p$ following different trends for the vanishing and non-vanishing $xy$-anisotropy. These constitute specific examples, other than the well-known examples of GHZ~\cite{greenberger1989} and GHZ-like states~\cite{hein2006,amaro2018}, and the case of persistence of localizable entanglement over a distance longer than the nearest-neighbor spin pair in a short-ranged 1D quantum spin model~\cite{verstraete2004},  where the partial trace-based and measurement-based entanglement behave differently. 

We further investigate  the situation where the single-qubit measurements being made on all central qubits to be noisy~\cite{Mondal2024}, assuming the presence of white noise in the measurement devices. We demonstrate that the partial trace-based bipartite peripheral entanglement and the LBPE appear as respectively the lower and the upper bounds of the localizable bipartite entanglement obtained via noisy measurements with specific noise strengths, where the lower  (upper) bound is obtained when the noise is maximum (minimum). We point out that in the case of the $n_0/n_p\ll 1$ limit of the spin-star system, localizable bipartite entanglement corresponding to all values of the noise strength, including the maximum and the minimum values, exhibit the same trends with periphery-size, $n_p$, as in the case of the noiseless scenario. However, in the case of the limit $n_0/n_p\rightarrow 1$, localizable bipartite entanglement does not exhibit a logarithmic trend except in the case of vanishing noise (i.e., perfect projection measurements).   

The rest of the paper is organized as follows. In Sec.~\ref{sec:star_network}, we introduce the XYZ model on the spin-star system, and discuss its diagonalization and  the features of the degeneracy of the ground state. In Sec.~\ref{sec:static_entanglement}, we provide brief definitions of the localizable bipartite peripheral entanglement in the case of unsharp measurements. We also compute the localizable bipartite entanglement on the peripheral spins using noiseless projection measurements (i.e., in the limit of vanishing noise) on the central spins, and discuss its properties in the three limits of the spin-star system. The features of the dynamics of localizable bipartite peripheral entanglement due to the introduction of a magnetic field of constant magnitude is also explored. Sec~\ref{sec:noisy_measurements} probes the localizable entanglement on the peripheral spins when the measurements are noisy, and the concluding remarks and the outlook are presented in Sec.~\ref{sec:conclude}.

\begin{figure*}
    \centering
    \includegraphics[width=.7\textwidth]{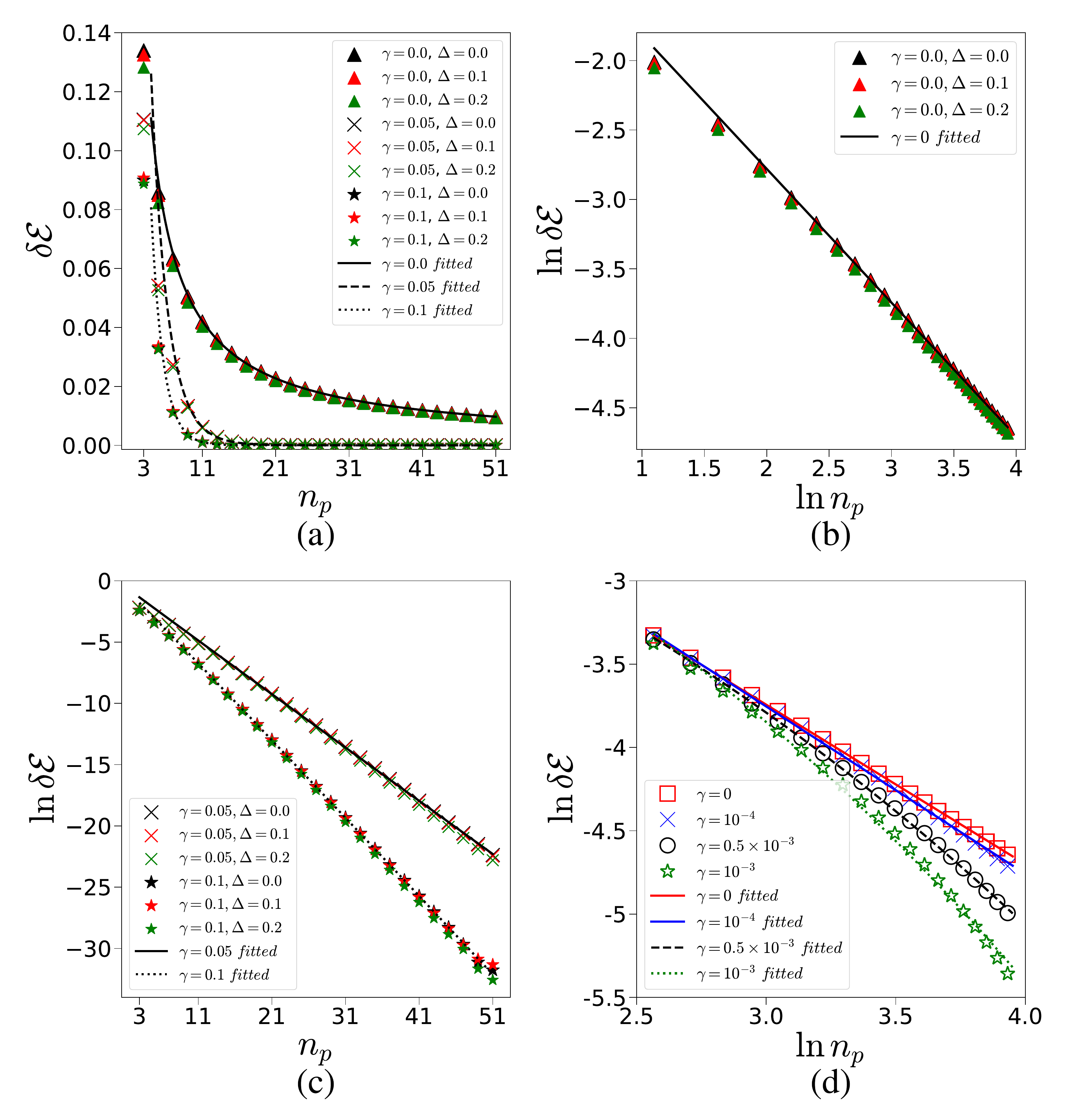}
    \caption{Variations of (a) $\delta \mathcal{E}$ with $n_p$, (b) $\ln \delta\mathcal{E}$ with $\ln n_p$ (for $\gamma=0$), and (c) $ \ln \delta\mathcal{E}$ with $n_p$ (for $\gamma\neq 0$) for different values of $\gamma$ and $\Delta$, when $n$ is even. The data is fit to Eq.~(\ref{eq:double_exponential}). For $\gamma=0$, the fitting parameters are $b=0.41(8)$ and $m=0.95(7)$. On the other hand, for $\gamma=0.05$ and $0.1$, the fitting parameters are $b=0.60(4)$, $m=0.88(7)$, and $b=0.64(3)$, $m=0.98(3)$. \JK{The variations of the energy gap for smaller values of $\gamma$ ($0\leq\gamma\leq 10^{-3}$) are shown in (d), where the $\gamma = 10^{-3}$ and $0.5\times 10^{-3}$ curve are fitted to the function corresponding to $\gamma\neq 0$ in Eq.~(\ref{eq:double_exponential}), with fitting parameters $b=-1.3(8)$ and $m=0.34(0)$ (for $\gamma=10^{-3}$), and $b=-1.58(7)$, $m=0.29(0)$ (for $\gamma=0.5\times10^{-3}$), while data for $\gamma=10^{-4}$ are fitted to the polynomial corresponding to $\gamma=0$ in Eq.~(\ref{eq:double_exponential}) with fitting parameters $b=0.47(6)$ and $m=1.00(4)$.} All quantities plotted are dimensionless.}
    \label{fig:energy_gap}
\end{figure*}

\section{XYZ model on a star network of spins }
\label{sec:star_network}

We now set the notations for the system, and discuss its features that are going to be useful for analysing the entanglement properties of its ground state(s).  We consider a star-network~\cite{Hutton2004} of $n=n_0+n_p$ spin-$1/2$ particles each of which represents a qubit, where $n_0$ \emph{central qubits} are surrounded by $n_p$ \emph{peripheral qubits} such that each of the central qubits interacts with all of the peripheral qubits, but there is no interaction among the central qubits and the peripheral qubits themselves (see Fig.~\ref{fig:system}). From now on, we use the terms ``spin" and ``qubit" interchangeably.  Without any loss in generality, we label the central qubits as $0,1,2,\cdots,n_0-1$, and the $n_p$ peripheral qubits as $n_0,n_0+1,n_0+2,\cdots,n_0+n_p-1$. We consider three limits of the model, 
\begin{enumerate} 
\item[(a)] the \emph{large periphery limit}, $n_0/n_p\ll 1$, where the size of the periphery is far larger than the size of the center, 
\item[(b)] the \emph{large center limit}, $n_0/n_p\gg 1$, where the central qubits outnumber the peripheral qubits by a large number,  and 
\item[(c)] the \emph{competing center limit}, $n_0/n_p\rightarrow 1$, where the number of central qubits is comparable to the number of peripheral qubits.  
\end{enumerate}
See Fig. \ref{fig:system} for pictorial representations of these limits of the system. We assume each of the central qubits $j$ interacting with all of the peripheral qubits via the interaction Hamiltonian~\cite{korepin_bogoliubov_izergin_1993} 
\begin{eqnarray}
H_{s}&=& K\sum_{i=0}^{n_0-1}\sum_{j=n_0}^{n_0+n_p-1}\left[(1+\gamma)S^x_i S^x_{j}+(1-\gamma) S^y_i S^y_{j}\right]+\Delta K\sum_{i=0}^{n_0-1}\sum_{j=n_0}^{n_0+n_p-1}S^z_i S^z_{j},
\label{eq:star-ham}
\end{eqnarray} 
Here,  $i$ and $j$ are the qubit-indices corresponding to the central and the peripheral qubits respectively, $S^{\alpha}_i=\sigma^{\alpha}_i/2$  is the standard representation of spin operators corresponding to the lattice site $i$, with $\sigma^{\alpha}$ being the Pauli matrices, $\alpha=x,y,z$,  and $K$ is the strength of the exchange interaction. The dimensionless parameters $\gamma$ and $\Delta$ respectively represent the $xy$- and the $z$-anisotropy corresponding to all pairs of qubits, with $-1\leq\gamma\leq1$ and $-1\leq\Delta\leq1$. The Hamiltonian $H_s$ represents a number of paradigmatic quantum spin Hamiltonians on the star network for different values of $\gamma$ and $\Delta$, including the XY model ($0<|\gamma|<1$, $\Delta=0$), the XX model ($\gamma=0,\Delta=0$) the classical Ising model ($\gamma=\pm 1$, $\Delta=0$), the isotropic Heisenberg model $(\gamma=0,\Delta=1)$, and the Heisenberg model  with a $z$-anisotropy  ($\gamma=0$, $0<|\Delta|<1$). In this paper, we consider the most general XYZ model  ($0< |\gamma|< 1$, $0 < |\Delta| < 1$) on the spin-star system in all three limits. 

\subsection{Large periphery limit}
\label{subsec:large_periphery_energy}

We first consider  the large periphery limit, and choose the case of a single central qubit ($n_0=1$), and increase $n_p$ in order to attain this limit. We identify $|K|$ as the natural energy scale of the system, and write the dimensionless spin-star Hamiltonian   
$H_s/|K|\rightarrow H_s$  as 
\begin{eqnarray}
\label{eq:star-xyz_ordered} 
H_s &=& \pm\frac{1}{2}\left\{S_0^+(J_p^- + \gamma J_p^+) + S_0^-(J_p^+ + \gamma J_p^-)\right\}+ \Delta S_{0}^{z} J_{p}^{z},\nonumber\\ 
\end{eqnarray}
where  
$J_p^\alpha=\sum_{i=1}^{n_p}S^\alpha_{i}$, $\alpha=x,y,z$, are defined on the Hilbert space of the peripheral qubits, and the signs of $H_s$ is determined by whether the spin-spin interaction is ferromagnetic (FM, $K>0$) or antiferromagnetic (AFM, $K<0$).  
Note that the operators $J_p^{\alpha}$, $\alpha=x,y,z$, obeys the usual commutation and anticommutation relation of angular momentum operators, implying that the peripheral qubits collectively behave as a spin-$n_p/2$ particle with the spin operator $\mathbf{J}_p=(J_p^x,J_p^y,J_p^z)$, while the total spin operator for the star network is given by $\mathbf{S}_0+\mathbf{J}_p$, with $\mathbf{S}_0=(S^x_0,S^y_0,S^z_0)$. We also define $S^\pm_i=S^x_i\pm\text{i}S^y_i$ for the $i$th spin, and subsequently $J^\pm_p=\sum_{i=1}^{n_p} S^\pm_i$.

\subsubsection{Diagonalization}
\label{subsubsec:diagonalization}

Since $[H_s,\mathbf{J}_p^2]=[H_s,\mathbf{S}_0^2]=0$, $H_s$ is block diagonal in the basis $\ket{b}=\ket{S_0,m_0}\otimes\ket{J_p,m_p}$, where
\begin{eqnarray}
\mathbf{S}_0^2\ket{S_0,m_0}&=&S_0 (S_0+1)\ket{S_0,m_0},\nonumber\\
\mathbf{J}_p^2\ket{J_p,m_p}&=&J_p (J_p+1)\ket{J_p,m_p}.
\end{eqnarray} 
Noticing that  
\begin{enumerate}
\item[(a)] $S_0$ only has one allowed value $S_0=\frac{1}{2}$ implying $m_0=\pm 1/2$, and
\item[(b)] different blocks of $H_s$ corresponding to a fixed value of $m_0$ can be labelled by specific values of $J_p$, where $m_p$ in each block can take $2J_p+1$ values, $-J_p\leq m_p\leq J_p$, 
\end{enumerate} 
for each block, it is sufficient to represent the basis as $\ket{b}=\ket{m_0,m_p}=\ket{m_0}\otimes\ket{m_p}$. Evidently, the ground state of $H_s$ belongs to one of these blocks with a specific value of $J_p$. Our numerical analysis suggests that irrespective of the system size $n$, the ground state $\ket{\Psi_0}$  always belongs to the sector with $J_p=n_p/2$ (cf.~\cite{Richter_1994}, and also ~\cite{Latorre2005} for a similar property in the LMG model), while the same is true for the first excited state $\ket{\Psi_1}$ for $n_p\geq 3$ (see Appendix~\ref{app:small_system} for a demonstration with $n_p=2,3$).

Let us now set $\ket{m_i=\pm 1/2}$ basis as the computational basis $\{\ket{0}\equiv\ket{1/2},\ket{1}\equiv\ket{-1/2}\}$ for the qubit $i$, such that 
\begin{eqnarray}
    S^z_i\ket{+1/2}&=&(+1/2)\ket{+1/2},\nonumber\\
    S^z_i\ket{-1/2}&=&(-1/2)\ket{-1/2}.
\end{eqnarray}
Using this, we write the permutation-invariant Dicke state~\cite{dicke1954} on the $n_p$  peripheral spins with $n_p-l$ excitations:
\begin{eqnarray}
    \ket{D^{n_p}_l} &=& \frac{1}{\sqrt{\genfrac(){0pt}{1}{n_p}{l}}}\sum_i\mathcal{P}_i\left(\ket{1}^{\otimes l}\ket{0}^{\otimes n_p-l}\right)
\end{eqnarray}
with the summation being over all possible permutations $\mathcal{P}_i$ over $n_p$-spin product states, where $l$ spins are in the ground state $\ket{1}$, and the rest $n_p-l$ spins are in the excited state $\ket{0}$, $0\leq l\leq n_p$ (note that we write $\sigma^z=\ket{0}\bra{0}-\ket{1}\bra{1}$). Noting that 
\begin{eqnarray}
    J_p^z\ket{D_l^{n_p}}&=&\lambda_l\ket{D_l^{n_p}}, \\ J_p^2\ket{D_l^{n_p}}&=&\lambda\ket{D_l^{n_p}},\\
    J_p^\pm\ket{D_l^{n_p}}&=&\Big[\lambda-\lambda_l\left(\lambda_l \pm 1 \right)\Big]^{\frac{1}{2}}\ket{D_{l\mp 1}^{n_p}},
\end{eqnarray}
where $\lambda_l=\frac{n_p}{2}-l$ and $\lambda=\frac{n_p}{2}\left(\frac{n_p}{2}+1\right)$, we further identify
\begin{eqnarray}
\ket{D_l^{n_p}}\equiv\left|J_p=\frac{n_p}{2},m_p=\frac{n_p}{2}-l\right\rangle;\quad 0\leq l\leq n_p.
\end{eqnarray}
Therefore, the  $J_p=n_p/2$  block of $H_s$, which we denote by $\mathcal{H}_{n_p/2}$, is a $2(n_p+1)$-dimensional subspace spanned by the permutation-invariant $n_p$ qubit Dicke states, with
\begin{eqnarray}
    \ket{m_0,m_p}\equiv\ket{m_0}\otimes\ket{D^{n_p}_l};\quad 0\leq l\leq n_p,
\end{eqnarray}
and $\mathcal{H}_{n_p/2}$ takes the form
\begin{eqnarray}
    \mathcal{H}_{n_p/2}=\begin{bmatrix}
    B & A \\ A^T & B^\prime
    \end{bmatrix},
    \label{eq:n_2_block}
\end{eqnarray}
with the matrix elements for the $n_p+1$ dimensional matrices $A$ and $B$ as  
\begin{eqnarray}\label{eq:A}
A_{i,j} &=& \left[\lambda- \lambda_j(\lambda_j-1)\right]^{\frac{1}{2}} \delta_{i-1,j}/2  + \gamma \left[\lambda- \lambda_j(\lambda_j+1)\right]^{\frac{1}{2}} \delta_{i+1,j}/2, \\
\label{eq:B}
B_{i,j}&=& \Delta\lambda_i\delta_{i,j}/2 = -B_{i,j}^\prime,
\end{eqnarray}
where $i,j\in [0,n_p]$. Due to the linear increase in dimension with increasing $n_p$, $\mathcal{H}_{n_p/2}$ can be numerically diagonalized to obtain the ground state of the model even for large $n_p$.

We further re-arrange the basis corresponding to the subspace hosting $\mathcal{H}_{n_p/2}$ such that $\mathcal{H}_{n_p/2}$ takes the form 
\begin{eqnarray}
    \mathcal{H}_{n_p/2} &=&
    \begin{bmatrix}
       A_1 & 0 \\ 0 & A_2
    \end{bmatrix},
    \label{eq:block_diagonal_forms}
\end{eqnarray}
where each of $A_1,A_2$ are $(n_p+1)$-dimensional matrices, and the basis has been grouped as \small 
\begin{eqnarray}
    A_1 &:& \{\ket{+1/2}\ket{-n_p/2+2k},\ket{-1/2}\ket{-n_p/2+2k+1}\},\nonumber\\
    A_2 &:& \{\ket{+1/2}\ket{-n_p/2+2k+1},\ket{-1/2}\ket{-n_p/2+2k}\},   
    \label{eq:grouping}
\end{eqnarray} \normalsize 
with $0\leq k\leq n_p/2$ for even $n_p$, and $0\leq k\leq (n_p-1)/2$ when $n_p$ is odd. The utility of this re-arrangement will be clear in subsequent discussions. 

\subsubsection{Ground state degeneracy}
\label{subsubsec:single_central_spin_ground_state}

Let us now define 
\begin{eqnarray}
    \mathcal{O}=S_0^x\bigotimes_{i=1}^{n_p}S_i^x, 
\end{eqnarray} 
where $[H_s,\mathcal{O}]=0$. We first focus on the case of even $n_p$, and notice that $\mathcal{O}$ connects the basis elements corresponding to $A_1$ and that corresponding to $A_2$. Let us denote the eigenstate corresponding to the ground state energy $\mathcal{E}_0$ coming from $A_1$ is $\ket{\psi_0}_{A_1}$, and the same for $A_2$ is $\ket{\psi_0}_{A_2}$, while $\ket{\psi_0}_{A_1}=\mathcal{O}\ket{\psi_0}_{A_2}$, and vice-versa. The doubly-degenerate (DD) ground state of the system in the common eigenspace of $\mathcal{O}$ and $H_s$ is given by (see Appendix~\ref{app:small_system} for an example) 
\begin{eqnarray}
    \ket{\Psi_0^{\pm}}=\frac{1}{\sqrt{2}}\left(\ket{\psi_0}_{A_1}\pm\ket{\psi_0}_{A_2}\right). 
    \label{eq:degenerate_ground_states}
\end{eqnarray}
The double degeneracy in the ground state of $H_s$ is found for all the allowed values of $0\leq|\gamma|\leq 1$, and $0\leq |\Delta|< 1$ when $n_p$ is even (i.e., when $n=n_p+1$ is odd).

\begin{figure}
    \centering
    \includegraphics[width=0.5\linewidth]{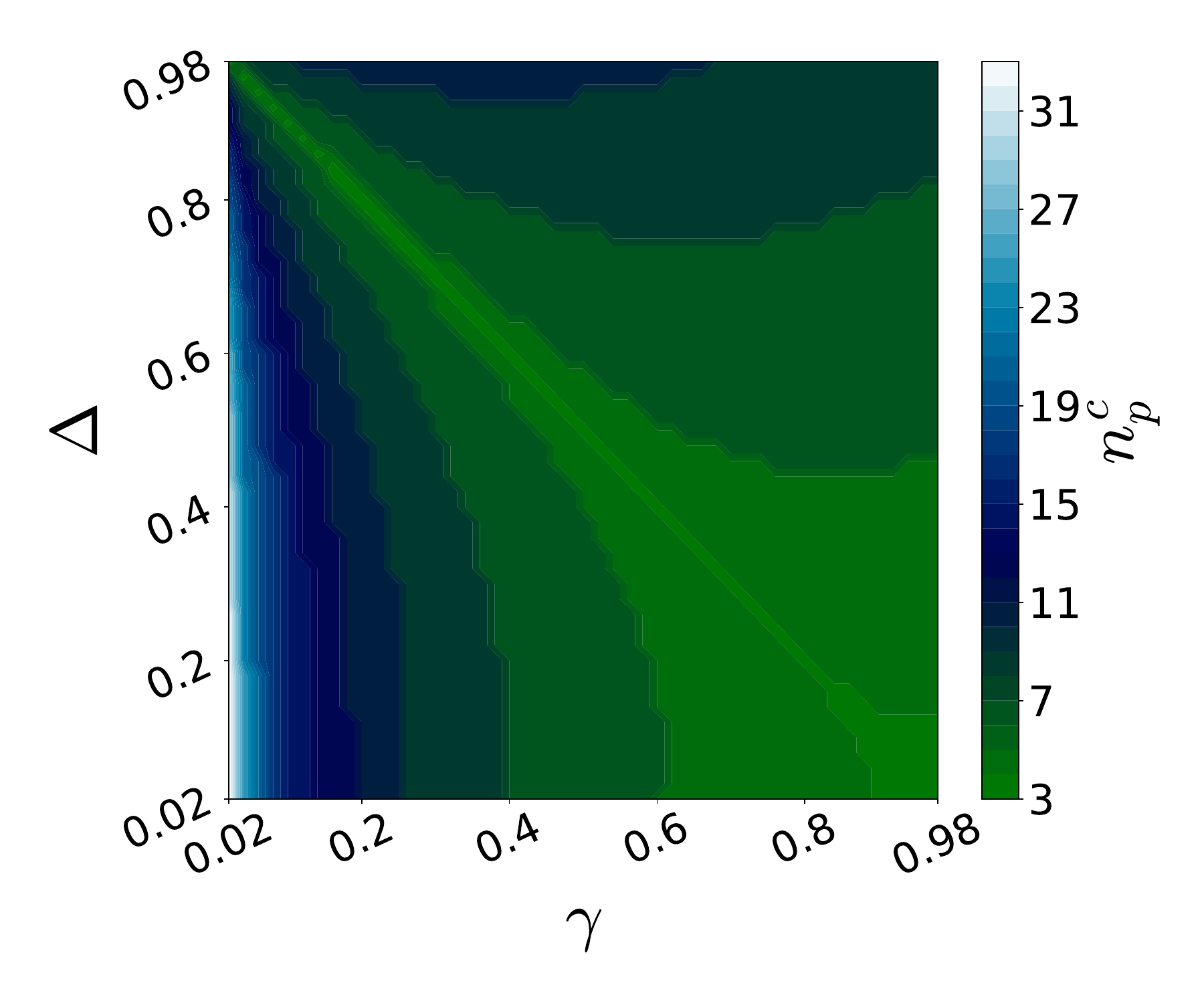}
    \caption{Variation of $n_p^c$ as a function of $\gamma$ and $\Delta$ in the ranges $0<\gamma,\Delta<1$. All quantities plotted are dimensionless.}
    \label{fig:nc}
\end{figure}

On the other hand, in the case of odd $n_p$ (i.e., for even $n=n_p+1$), application  of $\mathcal{O}$ on the eigenstates of $A_1$ ($A_2$) does not take the state out of the subspace of $A_1$ ($A_2$). Apart from the lines $|\Delta|=1$ in the $(\gamma,\Delta)$ parameter space where the ground state is DD (see Appendix~\ref{app:small_system} for examples), at all other allowed values of the anisotropy parameters ($0\leq|\gamma|< 1$ (at $\gamma=1$, ground states in the cases of both odd and even $n_p$ are doubly degenerate) and $0\leq|\Delta|<1$), the ground state in non-degenerate (ND). However, the energy gap $\delta \mathcal{E}=\mathcal{E}_1-\mathcal{E}_0$ of the system, $\mathcal{E}_1$ $(\mathcal{E}_0)$ being the energy corresponding to the first excited (ground) state, decreases with increasing $n_p$ as (see Fig.~\ref{fig:energy_gap})
\begin{eqnarray}
    \delta \mathcal{E}&\sim& \left\{\begin{array}{cc}
    bn_p^{-m} & \text{for }\gamma=0 \\
    \exp\left(-bn_p^{m}\right) & \text{for }\gamma\neq 0
    \end{array}\right.
    \label{eq:double_exponential}
\end{eqnarray}
where the parameters $b$ and $m$ can be obtained by fitting numerical data (note that one can also include an additive constant $a$ in Eq.~(\ref{eq:double_exponential}) similar to Eq.~(\ref{eq:energy_fit_np}), which vanishes). For $\gamma=0$, $\delta\mathcal{E}$ decreases slowly, and \emph{vanishes} when the periphery-size $n_p$ is beyond a critical value $n_p^c\geq 5\times 10^3$ (i.e., when the system size is beyond a critical size $n_c=1+n_p^c$). This critical periphery-size, $n_p^c$, is a function of the chosen system parameters $(\gamma,\Delta)$.  The collapse of $\delta \mathcal{E}\rightarrow 0$ becomes more  rapid as the value of $\gamma$ increases, and is achieved for $n_p^c\geq 13$, when $\gamma\geq 10^{-1}$, while the effect of $\Delta$ on $n_p^c$ for a fixed $\gamma$ is nominal (see Fig.~\ref{fig:nc}). For $n_p\geq n_p^c$ with odd $n_p$, the ground states of the system, denoted by $\ket{\Psi_0^{\pm}}$ for consistency with energy eigenvalues $\mathcal{E}_1$ and $\mathcal{E}_0$ respectively, can be considered to be \emph{effectively}  DD (EDD). \JK{We point out here that the transition to the polynomial decay of the energy gap ($\gamma=0$) from the exponential decay ($\gamma\neq 0$) is a gradual one that takes place with decreasing $\gamma$. As also indicated by our data, for sufficiently small ($\lesssim 10^{-4}$) anisotropy parameter $\gamma$, a polynomial decay fits the variation of the energy gap with $n_p$ better, which is in agreement of our definition of zero for $\delta\mathcal{E}$ (see Fig.~\ref{fig:energy_gap}(d)).} 

\JK{A word on the definition of the \emph{vanishing} of $\delta\mathcal{E}$ is in order here. In our numerical analysis, we assume $\delta\mathcal{E}=0$ when $\delta\mathcal{E}<\delta\mathcal{E}_c$, and the choice of the \emph{cut-off} $\delta\mathcal{E}_c$, indeed, depends on our numerical precision. In this paper, we report all data up to at most the third decimal place, and the data for the critical periphery-size presented in the Fig.~\ref{fig:nc} corresponds to $\delta\mathcal{E}_c=10^{-4}$. We have performed a detailed numerical analysis on the dependence of the critical periphery-size on the chosen cut-off, and our investigation suggests a slow change (increase) of the critical periphery size with decreasing $\delta\mathcal{E}_c$, leading to no qualitative, or drastic quantitative change in the results with a variation in the chosen cut-off.}

Noticing the invariance of $H_s$ under permutation of qubits on the periphery irrespective of the value of $n_p$, the general form of the ground state $\ket{\Psi_0}$ (degenerate, or non-degenerate), obtained from diagonalizing $\mathcal{H}_{n_p/2}$, can be written as 
\begin{eqnarray}
    \ket{\Psi_0}&=&\sum_{l=0}^{n_p} c_{l}\ket{\psi_0}\otimes\ket{D^{n_p}_l}+\sum_{l^\prime=0}^{n_p} d_{l^\prime}\ket{\psi_0^\perp}\otimes\ket{D^{n_p}_{l^\prime}},
    \label{eq:ground_state_form}
\end{eqnarray}
where $\sum_{l=0}^{n_p}|c_l|^2+\sum_{l^\prime=0}^{n_p}|d_{l^\prime}|^2=1$, and $\{\ket{\psi_0},\ket{\psi_0^\perp}\}$ constitute a complete orthonormal basis in the Hilbert space of the central qubit. The coefficients $\{c_l\}$ and $\{d_{l^\prime}\}$ can be obtained by diagonalizing $\mathcal{H}_{n_p/2}$. If the ground state is doubly degenerate,  one works with the \emph{thermal ground state} (TGS) of $H_s$~\cite{osborne2002}, given by an equal mixture of the degenerate ground states $\ket{\Psi_0^{\pm}}$ as 
\begin{eqnarray}
    \rho_0=\frac{1}{2}\left(\ket{\Psi_0^+}\bra{\Psi_0^+}+\ket{\Psi_0^-}\bra{\Psi_0^-}\right). \label{eq:thermal_ground_state}
\end{eqnarray}
In the case of EDD ground states occurring for large odd $n_p$, one can also work with $\rho_0$, and consider it as the \emph{effective} TGS (ETGS), as there is negligible difference between $\mathcal{E}_0$ and $\mathcal{E}_1$.

\subsection{Large and competing center limit}

\begin{figure}
    \centering
    \includegraphics[width=0.45\linewidth]{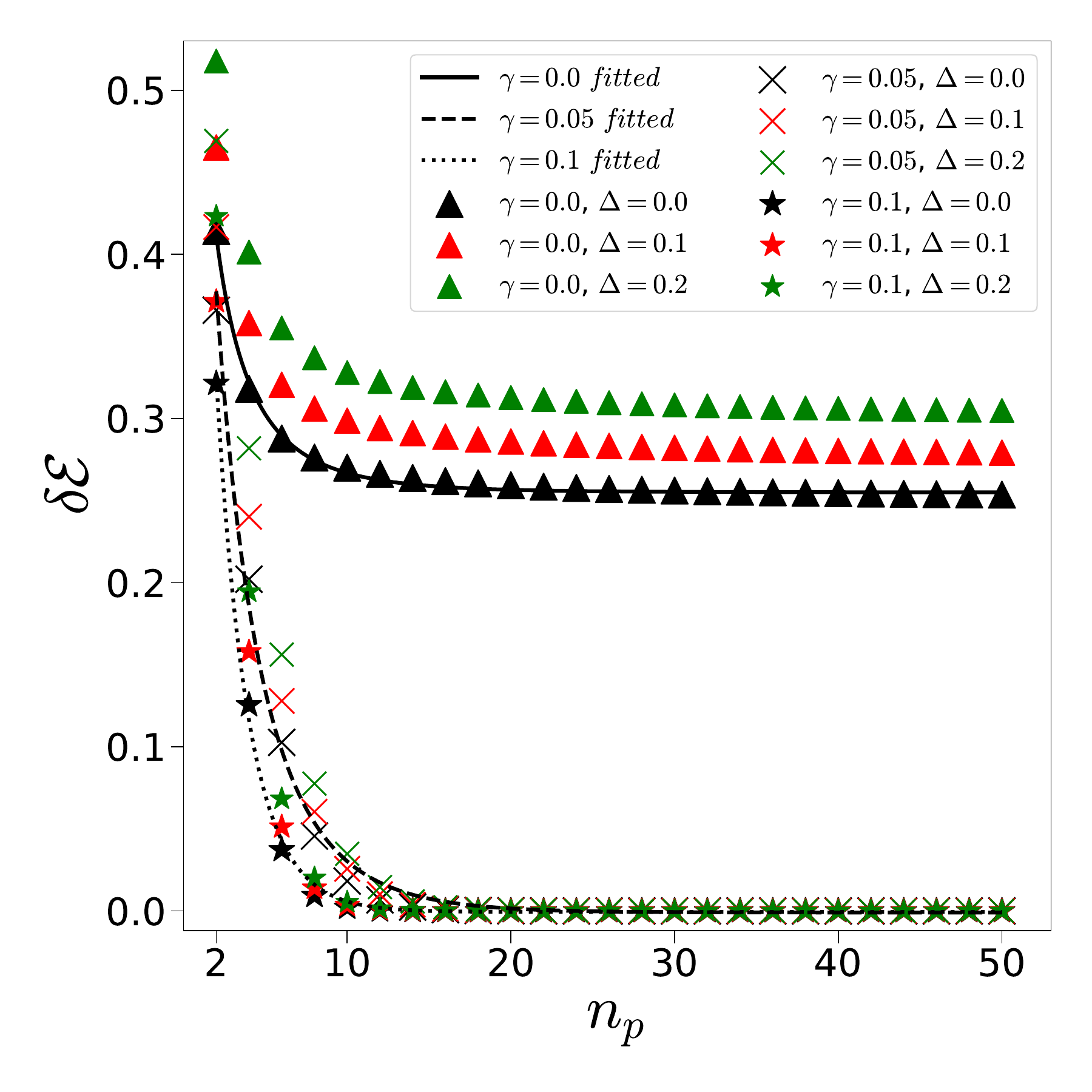}
    \caption{Variation of $\delta \mathcal{E}$ with $n_p$ in the limit $n_0/n_p\rightarrow 1$ where we fix $n_0=n_p$. The data corresponding to $\Delta=0$ is fitted to Eq.~(\ref{eq:energy_fit_np}) with fitting parameters ($a=0.255, b=1.2(7), m=0.5(4)$) for $\gamma=0$, ($a=0, b=0.5(6), m=0.7(9)$) for $\gamma=0.05$ and ($a=0, b=0.58(7), m=0.9(3)$) for $\gamma=0.1$.}
    \label{fig:energy_gap_n0=np}
\end{figure}

We now consider the XYZ model on the star network of spins when (a) the size of the center is far larger than the size of the periphery, i.e., $n_0/n_p\gg 1$, and when (b) $n_0/n_p\rightarrow 1$. Note that the former case is identical to the $n_0/n_p\ll 1$ limit of the model through an interchange of the central spins with peripheral spins, thereby exhibiting similar features of ground state energy as the limit $n_0/n_p\ll 1$. However, in the latter case,  the dimensionless spin-star Hamiltonian is given by (see Eq.~(\ref{eq:star-xyz_ordered}))  
\begin{eqnarray}
    H_s&=&\pm\frac{1}{2}\left[J_0^+(J_p^- + \gamma J_p^+) + J_0^-(J_p^+ + \gamma J_p^-)\right]+\Delta   J_0^z J_p^z,
\label{eq:Hamiltonian_coupled_central}
\end{eqnarray}
where $J_0^{\pm}=\sum_{i=0}^{n_0-1}S_i^{\pm}$, $J_0^{z}=\sum_{i=0}^{n_0-1}S_i^{z}$, $J_p^{\pm}=\sum_{i=n_0}^{n_0+n_p-1}S_i^{\pm}$, and $J_p^{z}=\sum_{i=n_0}^{n_0+n_p-1}S_i^{z}$. Analysis for the Hamiltonian (\ref{eq:Hamiltonian_coupled_central}) is similar to that of the Hamiltonian (\ref{eq:star-xyz_ordered}) in the case of the single central spin, and the values of $J_0$ and $J_p$ corresponding to the ground and the first excited states of $H_s$ are $n_0/2$ and $n_p/2$ respectively, independent of the system parameters $\gamma$, and $\Delta$. Upon diagonalization, doubly degenerate ground states are found when $n=n_0+n_p$ is odd irrespective  of the values of $(\gamma,\Delta)$, while the effective degeneracy in ground states for large system size is observed when $n$ is even only for $\gamma\neq 0$ (see Fig.~\ref{fig:energy_gap_n0=np}). Note that in the present case, $\delta \mathcal{E}$ approaches saturation (for $\gamma=0$) or vanishing (for $\gamma\neq 0$) as 
\begin{eqnarray}
\delta\mathcal{E}\sim a + \exp\left(-bn_p^{m}\right), 
\label{eq:energy_fit_np}
\end{eqnarray}
where in contrast to Eq.~(\ref{eq:double_exponential}), the values of $|a|\neq 0$. 
For $\gamma=0$, according to our numerical investigation up to $n_p=n_0=50$, $\delta\mathcal{E}$ saturates to a non-zero value, which increases with an increase in the value of $\Delta$, implying a ND ground state for $\gamma=0$ up to the maximum system-size investigated in this paper.

\section{Localizable bipartite entanglement on the periphery}
\label{sec:static_entanglement}

In this section, we set up the mathematical premises for \emph{localizable entanglement} (LE)~\cite{verstraete2004,sadhukhan2017,Mondal2024} assuming that the single-qubit measurements can, in principle, be imperfect (noisy)~\cite{Mondal2024}. We also discuss the features of the entanglement \emph{localized} over the periphery of the spin-star system assuming the measurements to be \emph{perfect}~\cite{verstraete2004,sadhukhan2017}.

\subsection{Localizing entanglement}
\label{subsec:localizable_entanglement}

Let us consider a bipartition $A:B$ of a $n$-qubit system described by the state $\rho$, where we aim to quantify entanglement in $B$. The size of the individual partitions $A$ and $B$ are respectively determined by the number of qubits $n_A$ and $n_B$ in them ($n=n_A+n_B$). 
We assume the subsystem $A$ to be the measured subsystem, where \emph{independent} \emph{unsharp} measurements, with corresponding POVM element 
\begin{eqnarray}
    P(\lambda_a,\vec{\eta}_a)=\frac{1}{2}\left[I_a+\lambda_a\vec{\eta}_a.\vec{\sigma}_a\right]
    \label{eq:povm}
\end{eqnarray}
are performed on each qubits $a\in A$. Here, we have assumed white noise~\cite{Mondal2024,nielsen2010} on the measurement apparatus, with $I_a$ being the identity operator in the qubit Hilbert space, $\vec{\eta}_a=\eta_a\hat{k}_a$ with $\eta$ $(0\leq \eta_a\leq 1)$ being the strength of the noise, $\lambda_a=\pm 1$ are the measurement outcomes, and $\hat{k}_a$ is the unit vector determining the measurement direction. We assume the measurement apparatus to be \emph{equally faulty} for all qubits in $A$, and fix $\eta_a=\eta\forall a\in A$, such that  Eq.~(\ref{eq:povm}) becomes 
\begin{eqnarray}
    P(\eta,\lambda_a,\hat{k}_a)=\frac{1}{2}\left[I_a+\eta\lambda_a\hat{k}_a.\vec{\sigma}_a\right].
    \label{eq:povm_simple}
\end{eqnarray}
After the measurement on all qubits in $A$ along the direction $\hat{k}\equiv\hat{k}_1\hat{k}_2\cdots\hat{k}_{n_A}$ and upon obtaining the measurement outcome $\lambda\equiv\lambda_1\lambda_2\cdots\lambda_{n_A}$,  the $n$-qubit system is described by the state 
\begin{eqnarray}
\varrho(\eta,\lambda,\hat{k})=\frac{M(\eta,\lambda,\hat{k})\rho M^\dagger(\eta,\lambda,\hat{k})}{p(\eta,\lambda,\hat{k})},
    \label{eq:post_measured_state}
\end{eqnarray}
which occurs with the probability 
\begin{eqnarray}
    p(\eta,\lambda,\hat{k})=\text{Tr}\left[M(\eta,\lambda,\hat{k})\rho M^\dagger(\eta,\lambda,\hat{k})\right],
\end{eqnarray}
where $M(\eta,\lambda,\hat{k})=\otimes_{a\in A}M(\eta,\lambda_a,\hat{k}_a)$, and $M(\eta,\lambda_a,\hat{k}_a)$ is defined as $P(\eta,\lambda_a,\hat{k}_a)=M^\dagger(\eta,\lambda_a,\hat{k}_a)M(\eta,\lambda_a,\hat{k}_a)$. Therefore, the \emph{average entanglement} over the ensemble $\{p(\eta,\lambda,\hat{k}),\varrho_B(\eta,\lambda,\hat{k})\}$ of all possible post-measured states on $B$, with $\varrho_B(\eta,\lambda,\hat{k})=\text{Tr}_A\left[\rho(\eta,\lambda,\hat{k})\right]$, is given by~\cite{Mondal2024} 
\begin{eqnarray}
    \left\langle E_B(\eta,\hat{k})\right\rangle=\sum_\lambda p(\eta,\lambda,\hat{k}) E\left(\varrho_B(\eta,\lambda,\hat{k})\right),
    \label{eq:avg_ent}
\end{eqnarray}
with $E$ being either a bipartite, or a multipartite entanglement measure~\cite{horodecki2009}, to be computed on $B$ over the qubits $b=1,2,\cdots,n_B$ in $B$. In this paper, we are interested in bipartite entanglement over a bipartition $B_1:B_2$ of $B$, where we assume that the partition $B_1$ $(B_2)$ has $n_{B_1}$ $(n_{B_2})$ qubits ($n_{B_1}+n_{B_2}=n_B$), and denote the computed average entanglement by $\left\langle E_{B_1:B_2}(\eta,\hat{k})\right\rangle$.

In spherical polar coordinate~\cite{nielsen2010}, 
\begin{eqnarray}
    \hat{k}_a = \left(\sin{\theta_a}\cos{\phi_a}, \sin{\theta_a}\sin{\phi_a}, \cos{\theta_a}\right)\forall a\in A,
\end{eqnarray} 
with $\theta_a,\phi_a\in\mathbb{R}$,    $\theta_a\in[0,\pi]$, and $\phi_a\in[0,2\pi]$, allowing  one to write $M(\eta,\lambda_a, \hat{k}_a)$ as 
\begin{eqnarray}
    M(\eta,\lambda_a,\theta_a,\phi_a)=\sqrt{\frac{1+\lambda_a\eta}{2}}\ket{\lambda^+_a}\bra{\lambda^+_a}+\sqrt{\frac{1-\lambda_a\eta}{2}}\ket{\lambda^-_a}\bra{\lambda^-_a}
\end{eqnarray}
where  
\begin{eqnarray}
    \ket{\lambda_a^{+}}&=&\cos{\frac{\theta_a}{2}}\ket{0_a}+e^{i\phi_a}\sin{\frac{\theta_a}{2}}\ket{1_a}, \nonumber\\ 
    \ket{\lambda_a^{-}}&=&\sin{\frac{\theta_a}{2}}\ket{0_a}-e^{i\phi_a}\cos{\frac{\theta_a}{2}}\ket{1_a},
    \label{eq:parametrization_2}
\end{eqnarray}
correspond to $\lambda_a=\pm 1$ respectively, such that  $\sum_{\lambda_a}P(\eta,\lambda_a,\hat{k}_a)=I$ for fixed $\eta$ and $\hat{k}_a$ (i.e., fixed $\theta_a$ and $\phi_a$). For a fixed value of $\eta$, this parametrization reduces Eq.~(\ref{eq:avg_ent}) to 
\begin{eqnarray}
    \left\langle E_{B_1:B_2}(\eta,\Theta)\right\rangle=\sum_\lambda p(\eta,\lambda,\Theta) E_{B_1:B_2}(\eta,\lambda,\Theta),
    \label{eq:avg_ent_2}
\end{eqnarray}
where $\Theta$ is the set of $2n_A$ real parameters $\{\theta_a,\phi_a\}\forall a\in A$, and we have written $E\left(\varrho_B(\eta,\lambda,\Theta)\right)$ as $E_{B_1:B_2}(\eta,\lambda,\Theta)$. A maximization of $\left\langle E_{B_1:B_2}(\eta,\Theta)\right\rangle$ over $\Theta$ provides the \emph{localizable entanglement} over the bipartion $B_1:B_2$ of the subsystem $B$, given by~\cite{Mondal2024} 
\begin{eqnarray}
\left\langle E_{B_1:B_2}(\eta)\right\rangle = \max_{\Theta}\left\langle E_{B_1:B_2}(\eta,\Theta)\right\rangle.
\label{eq:LE}
\end{eqnarray}
Note that  for $\eta=1$, $P(\eta,\lambda_a,\hat{k}_a)$ corresponds to a sharp projection measurement on the qubit $a\in A$, and Eq.~(\ref{eq:LE}) becomes
\begin{eqnarray}
 \left\langle E_{B_1:B_2}\right\rangle = \max_{\Theta}\left\langle E_{B_1:B_2}(\Theta)\right\rangle,
 \label{eq:noiseless_measurement}
\end{eqnarray}
which is the case considered in~\cite{verstraete2004,sadhukhan2017}. Note further that for a qubit $a\in A$, noiseless projection measurements of $S^z$, $S^x$, and $S^y$ correspond to $(\theta_a=0,\phi_a=0)$, $(\theta_a=\pi/2,\phi_a=0)$, and $(\theta_a=\pi/2,\phi_a=\pi/2)$ respectively.

\begin{figure}
    \centering
    \includegraphics[width=0.5\linewidth]{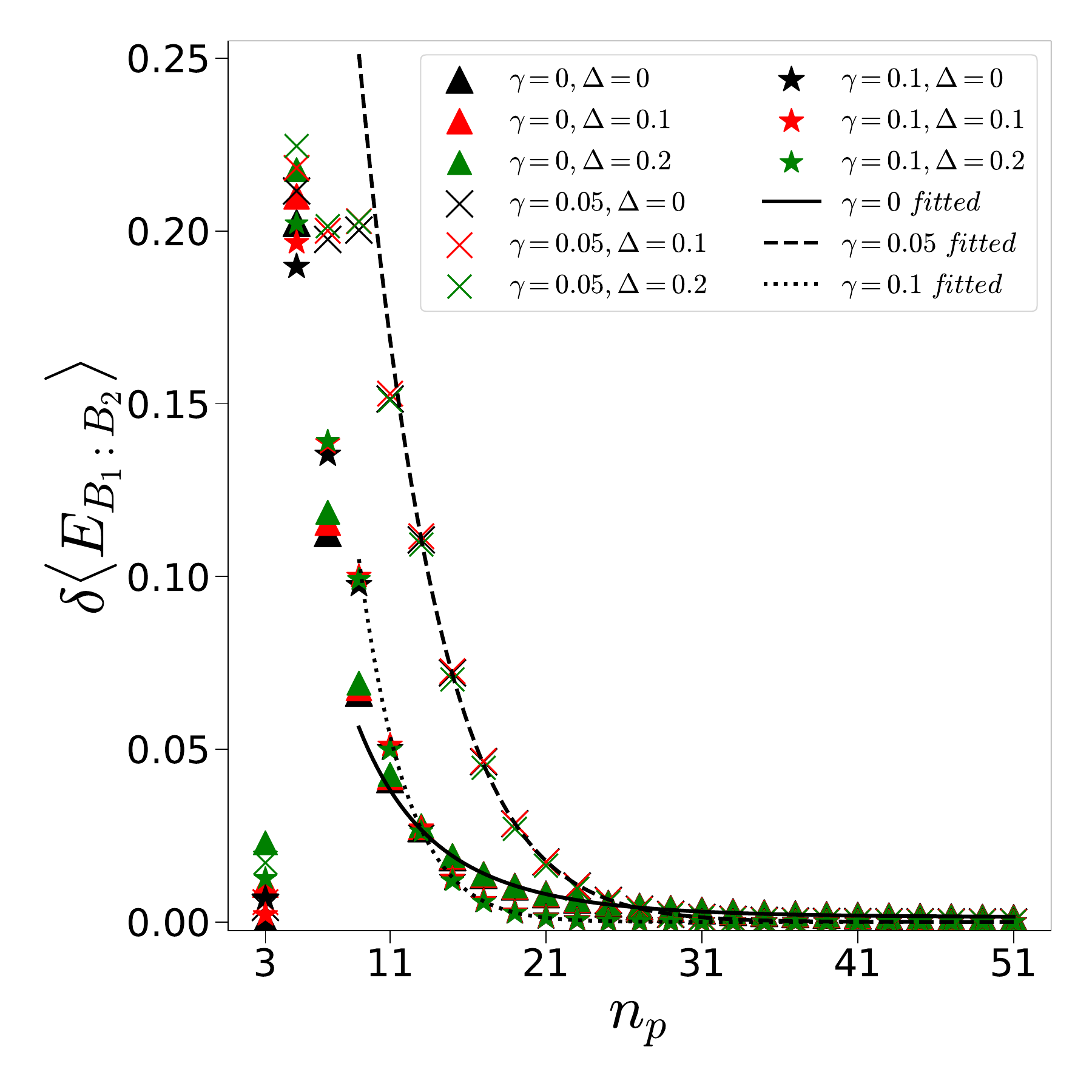}
    \caption{Variation of $\delta\langle E_{B_1:B_2}\rangle$ as function of $n_p$ for different values of $\gamma$ and $\Delta$, when $n$ is even. The data is fit to Eq.~(\ref{eq:entanglement gap fitt}). The fitting parameters are ($a=0.001$, $b=0.7(0)$, $m=0.6(4)$) for $\gamma=0$, ($a=0$, $b=0.08(5)$, $m=1.26(6)$)   for $\gamma=0.05$, and ($a=0$, $b=0.13(3)$, $m=1.28(8)$) for $\gamma=0.1$. All quantities plotted are dimensionless.}
    \label{fig:entanglement gap}
\end{figure}


To quantify bipartite entanglement over a partition $B_1:B_2$ in the state  $\varrho_B(\eta,\lambda,\Theta)$ on $B$, one needs to choose a bipartite entanglement measure $E$ that is computable for arbitrary values of $n_{B_1}$ and $n_{B_2}$. In this paper, we use \emph{logarithmic negativity}~\cite{plenio2005} for this purpose, which is defined as 
\begin{eqnarray}
   \mathcal{L} &=& \log_2(2\mathcal{N}+1),
\end{eqnarray}
where $\mathcal{N}$ is the \emph{negativity}~\cite{peres1996,vidal2002} of $\rho_B$, defined as 
\begin{eqnarray} 
\mathcal{N}&=&\frac{||\rho_{B}^{T_{B_1}}||-1}{2}=\left|\sum_{\lambda_i<0}\lambda_i\right|.
\end{eqnarray}
Here,  $\rho_{B}^{T_{B_1}}$ is the partially transposed density matrix $\rho_{B}$ with respect to the subsystem  $B_1$, $||.||$ denotes the trace norm, and $\lambda$ are the negative eigenvalues of $\rho_{B}^{T_{B_1}}$.

\subsection{Peripheral entanglement in large periphery limit: Logarithmic growth and anisotropy effect}
\label{subsec:static}

\begin{figure*}
    \centering
    \includegraphics[width=0.85\textwidth]{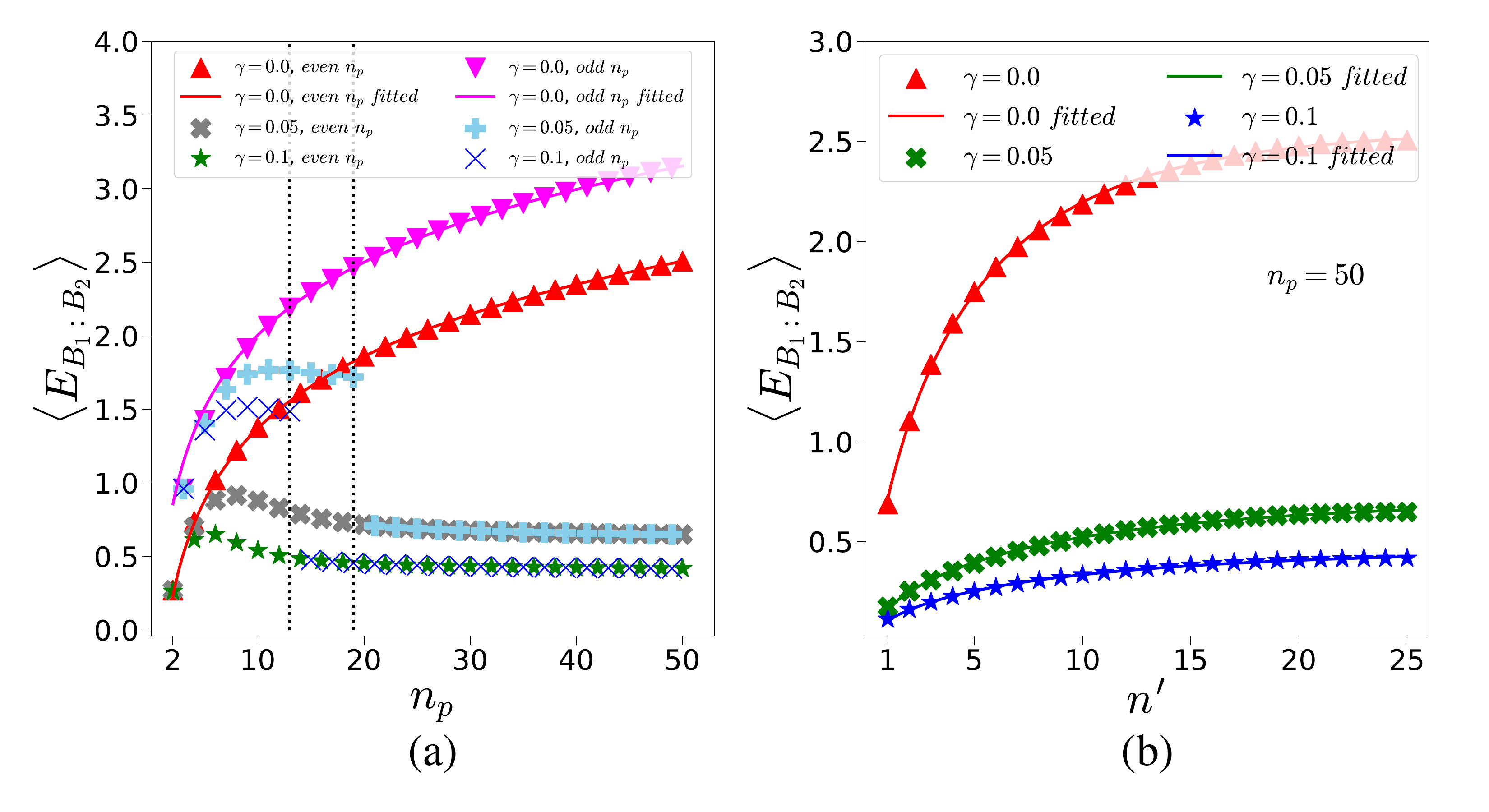}
    \caption{
    Variations of $\langle E_{B_1:B_2}\rangle$  as a function of (a) $n_{p}$ with $n^\prime=n_{p}/2$ \big($(n_{p}-1)/2$\big) when $n_{p}$ is even (odd), and as a function of (b) $n^\prime$ with $n_p=50$ (i.e., $n=51$, odd), for different values of $\gamma$, where $\Delta=0$.  In (a), the data for different values of $\gamma$ are fitted to Eq.~(\ref{eq:entanglement_fit_n}), while the  values of $n_p^c$ are $n_p^c=4999$ for $\gamma=0$ (not shown in the figure), $n_p^c=19$ for $\gamma=0.05$, and $n_p^c=13$ for $\gamma=0.1$. On the other hand, in (b), the data are fitted to Eq.~(\ref{eq:entanglement_fit_nprime}). In (a), the fitting parameters are  for ($a=-0.25(3)$, $b=0.489$) $\gamma=0$ and $n_p$ even, ($a=0.362$, $b=0.495$) for $\gamma=0$ and $n_p$ odd. In (b), the fitting parameters are ($a=2.56(5)$, $b=0.32(7)$, $m=0.77(3)$) for $\gamma=0.0$,  ($a=0.72(5)$, $b=0.26(1)$, $m=0.6(9)$) for $\gamma=0.05$, and  ($a=0.47(1)$, $b=0.25(8)$, $m=0.6(8)$ for $\gamma=0.1$. All quantities plotted are dimensionless.}  
    \label{fig:zero_field_dynamic}
\end{figure*}

We now consider the star-network of spins discussed in Sec.~\ref{sec:star_network}, and assume that the subsystem $A$ is constituted only of the  central spins, thereby having size $n_A=n_0$, while the peripheral spins constitute the partition $B$ of size $n_B = n_p$. We further assume partitions $B_1$ and $B_2$ being of the  sizes $n^\prime$ and $n_p-n^\prime$ respectively. For even $n_p$, $1\leq n^\prime \leq n_p/2$, while for odd $n_p$, $1\leq n^\prime\leq (n_p-1)/2$. Without any loss in generality, we assume that the spins $b=1,2,\cdots,n^\prime$ $\in B_1$, and $b=n^\prime+1,n^\prime+2,\cdots,n_p$ $\in B_2$.

We now compute the localizable bipartite peripheral entanglement (LBPE) $\langle E_{B_1:B_2}\rangle$, (assuming $\eta=1$, i.e., for noiseless projection measurements on all qubits in $A$, see Eq.~(\ref{eq:noiseless_measurement})) in $\ket{\Psi_0}$ (for ND ground state), or $\rho_0$ (for DD or EDD ground state). We first note that a measurement of $S^z_0$ on the central qubit  leads to  a post-measured ensemble $\{\rho_p^1,\rho_p^2\}$ on the peripheral qubits, with $\rho_p^{1,2}$ having the form 
\begin{eqnarray}
    \rho_p^{1,2}&=&\sum_{l,l^\prime=0}^{n_p}c_{l,l^\prime}\ket{D^{n_p}_l}\bra{D^{n_p}_{l^\prime}}.
    \label{eq:mixed_state_dicke_basis}
\end{eqnarray}
The advantage of the form of $\rho_p^{1,2}$ lies in the existence of the  prescription~\cite{Stockton2003} for computing the bipartite entanglement in a state of the form (\ref{eq:mixed_state_dicke_basis}) over arbitrary bipartition $B_1:B_2$, as  quantified by the logarithmic negativity, which further leads to the computation of LBPE corresponding to the $S^z_0$ measurement.
Indeed, the $S^z_0$ measurement on the central qubit, which corresponds to a  projection measurement on the central qubit in the basis $\{\ket{0},\ket{1}\}$, may not be optimal, and would only provide a lower bound corresponding to the actual localizable entanglement (see Eq.~(\ref{eq:LE}), where the optimization is effectively over all possible single-qubit Hermitian operators on all qubits in $A$). However, our numerical analysis for small systems ($n\leq 10$ with $n_0\leq 2$, $n_p\geq  8$) indicate that for $n_0\ll n_p$, $\{\ket{0},\ket{1}\}$ $S_0^z$ is the optimal basis for measuring  each qubit in $A$.

\paragraph{Variation with periphery-size.} Note that in the case of even $n_p$, the degenerate ground states $\ket{\Psi_0^\pm}$ are connected by the local unitary operator $\mathcal{O}$, and therefore have identical entanglement properties. However, this is not the case for odd $n_p$, and an interesting question would be how the two EDD states in the case of odd $n_p$  differ from each other in terms of values of LBPE, 
where $n^\prime$ is typically fixed at $n^\prime=n_p/2$ for even $n_p$ ($n^\prime=(n_p-1)/2$ for odd $n_p$), when $n_p$ is increased. In Fig.~\ref{fig:entanglement gap}, we plot $\delta \langle E_{B_1:B_2}\rangle=\left|\langle E_{B_1:B_2}^+\rangle-\langle E_{B_1:B_2}^-\rangle\right|$ 
as a function of $n_p$ for different values of $\gamma$ and  $\Delta$, where the $\pm$ in the superscripts in the expression for 
$\delta \langle E_{B_1:B_2}\rangle$ denote the states $\ket{\Psi_0^\pm}$   (see discussions below Eq.~(\ref{eq:double_exponential})) for which $\langle E_{B_1:B_2}\rangle$ is calculated. We find that $\delta\langle E_{B_1:B_2}\rangle$  
approaches zero as 
\begin{eqnarray}
\label{eq:entanglement gap fitt}
    \delta \langle E_{B_1:B_2}\rangle &\sim& a+ \exp\left(-b n_p^{m}\right),
\end{eqnarray}
similar to $\delta\mathcal{E}$ corresponding to the case of  $\gamma\neq 0$ (see
Eq.~(\ref{eq:double_exponential})). \JK{Note here that $\delta \langle E_{B_1:B_2}\rangle$, being the difference between the average entanglement in the periphery of the ground and the first excited states of the system, may, in principle, retain a non-zero value even when $n_p$ is large -- a possibility which is included in the numerical fit by retaining the parameter ``$a$".  However, our numerical investigation of  $\delta \langle E_{B_1:B_2}\rangle$ for the effectively doubly degenerate ground states in the case of odd $n_p$ reveals that $\delta \langle E_{B_1:B_2}\rangle\rightarrow 0$ asymptotically when $n_p$ increases. Therefore, fitting  $\delta \langle E_{B_1:B_2}\rangle$ to Eq.~(\ref{eq:entanglement gap fitt}) may lead to either zero, or a small a non-zero value of $a$ (see, for example, the case of $\gamma=0$ in Fig.~\ref{fig:entanglement gap}). In the latter case, the non-zero value decreases monotonically with increasing $n_p$, indicative of the asymptotic decay of $\delta \langle E_{B_1:B_2}\rangle$ with $n_p$.}

We now specifically discuss 
the case of 
$n_0=1$, and for demonstration, choose $n^\prime=n_p/2$ (for even $n_p$), and $(n_p-1)/2$ (for odd $n_p$). Fig.~\ref{fig:zero_field_dynamic} depicts the variation of $\langle E_{B_1:B_2}\rangle$ as a function of $n_p$ for different values of $\gamma$ and $\Delta$. Note that on the $(\gamma,\Delta)$ plane for odd $n_p$, the onset of $\delta\mathcal{E}=0$ (i.e., $n_p^c$, see Sec.~\ref{subsubsec:single_central_spin_ground_state}) is different for different points, which is indicated by the discontinuities in the variations of $\langle E_{B_1:B_2}\rangle$ for odd $n_p$, computed from the ND ground state $\ket{\Psi_0}$ (Eq.~(\ref{eq:ground_state_form})) for $n_p \leq n_p^c$,  and from the mixed ETGS $\rho_0$ (Eq.~(\ref{eq:thermal_ground_state})) for $n_p> n_p^c$. For $\gamma=0$, $\langle E_{B_1:B_2}\rangle$ exhibits a logarithmic dependence on $n_{p}$ as 
\begin{eqnarray}
    \langle E_{B_1:B_2}\rangle &\sim& a+b \log_2 n_p,
    \label{eq:entanglement_fit_n}
\end{eqnarray}
where the fitting parameters $a$ and $b$ can be obtained by fitting the data for $\langle E_{B_1:B_2}\rangle$ against $n_p$ (see Fig.~\ref{fig:zero_field_dynamic} for the example of $\Delta=0$) when $n_p>n_p^c$. Also our numerical analysis suggest that for $\gamma=0$, $a$ and $b$ depends on the choice of $\Delta$ only weakly for even $n_p$, and are independent of $\Delta$ for odd $n_p$.  However, in stark contrast, the logarithmic dependence is absent for $\gamma\neq 0$, where LBPE tends to saturate at a non-zero value for $n_0\ll n_p$. This result indicate a prominent change in the variations of LBPE with periphery-size for vanishing and non-vanishing anisotropy parameter $\gamma$ -- an effect that we refer to as the \emph{anisotropy effect}. Our analysis finds this feature to be unchanged for all constant ratios $n^\prime/n_p$
, where for the logarithmic dependence on $n_p$ in case of  $\gamma=0$, only values of $a$ and $b$ change. 

Note that for the center-periphery interaction described by the  XXZ model $(-1<\Delta<1,\gamma=0)$, the ground state of the spin-star Hamiltonian is doubly degenerate~\cite{Hutton2004,Deng_2008} for even $n_p$, having the form 
\begin{eqnarray}
\ket{\Psi_0}&=&\alpha\ket{\psi_0}\otimes\ket{D^{n_p}_{n_p/2}}+\beta \ket{\psi_0^\perp}\otimes\ket{D^{n_p}_{n_p/2-1}}, \nonumber\\
\ket{\Psi_0^\prime}&=&\alpha^\prime\ket{\psi_0}\otimes\ket{D^{n_p}_{n_p/2+1}}+\beta^\prime \ket{\psi_0^\perp}\otimes\ket{D^{n_p}_{n_p/2}},\label{eq:XXZ_lim_gstate_even}
\end{eqnarray}
with $|\alpha|^2+|\beta|^2=1$ and $|\alpha^\prime|^2+|\beta^\prime|^2=1$,  while for odd $n_p$, $H_s$ has a unique ground state, given by \small
\begin{eqnarray}
\ket{\Psi_0}&=&\frac{1}{\sqrt{2}}\left[\ket{\psi_0}\otimes\ket{D^{n_p}_{(n_p+1)/2}}+\ket{\psi_0^\perp}\otimes\ket{D^{n_p}_{(n_p-1)/2}}\right].\label{eq:XXZ_lim_gstate_odd}
\end{eqnarray}\normalsize 
Further, for the Dicke states $\ket{D^{n_p}_l}$ $\forall l$  in the range $1\leq l\leq n_p/2$, the entanglement over an equal bipartition $B_1:B_2$, as quantified by the von Neumann entropy~\cite{peres1996,vidal2002}, is shown to scale as
$\log_2(l)$~\cite{Stockton2003}. Since the post measured states on the peripheral qubits upon projective measurements on the central qubit in the states given in Eqs.~(\ref{eq:XXZ_lim_gstate_even}) and (\ref{eq:XXZ_lim_gstate_odd}) are Dicke states with $l\sim n_p$, and since the positivity of the partially transposed density matrix with respect to any one of the bipartitions is a necessary and sufficient condition for the separability of an arbitrary Dicke state~\cite{Quesada2017}, the results described in~\cite{Stockton2003} for the XXZ model are in alignment with our result using logarithmic negativity as the bipartite entanglement measure for the XYZ model with $\gamma=0$. On the other hand, for $\gamma \neq 0$, the ground state(s) of $H_s$ is of the form given in Eq.~(\ref{eq:ground_state_form}) with unnormalized post measured states on the peripheral qubits being superposition $\sum_l c_l \ket{D_l^{n_p}}$ in the Dicke basis with the coefficients $\{c_l\}$ being a function of $\gamma$ and $\Delta$. Note that while randomly generated symmetric states with the coefficients $\{c_l\}$ sampled from a Gaussian distribution exhibits a $\log_2(n_p)$ behaviour~\cite{Stockton2003}, symmetric states occurring in the XYZ model $(\gamma\neq 0)$ does not, as verified with our numerical analysis using both von Neumann entropy and logarithmic negativity, as demonstrated with the latter in Fig.~\ref{fig:zero_field_dynamic}.

\begin{figure*}
    \centering
    \includegraphics[width=\linewidth]{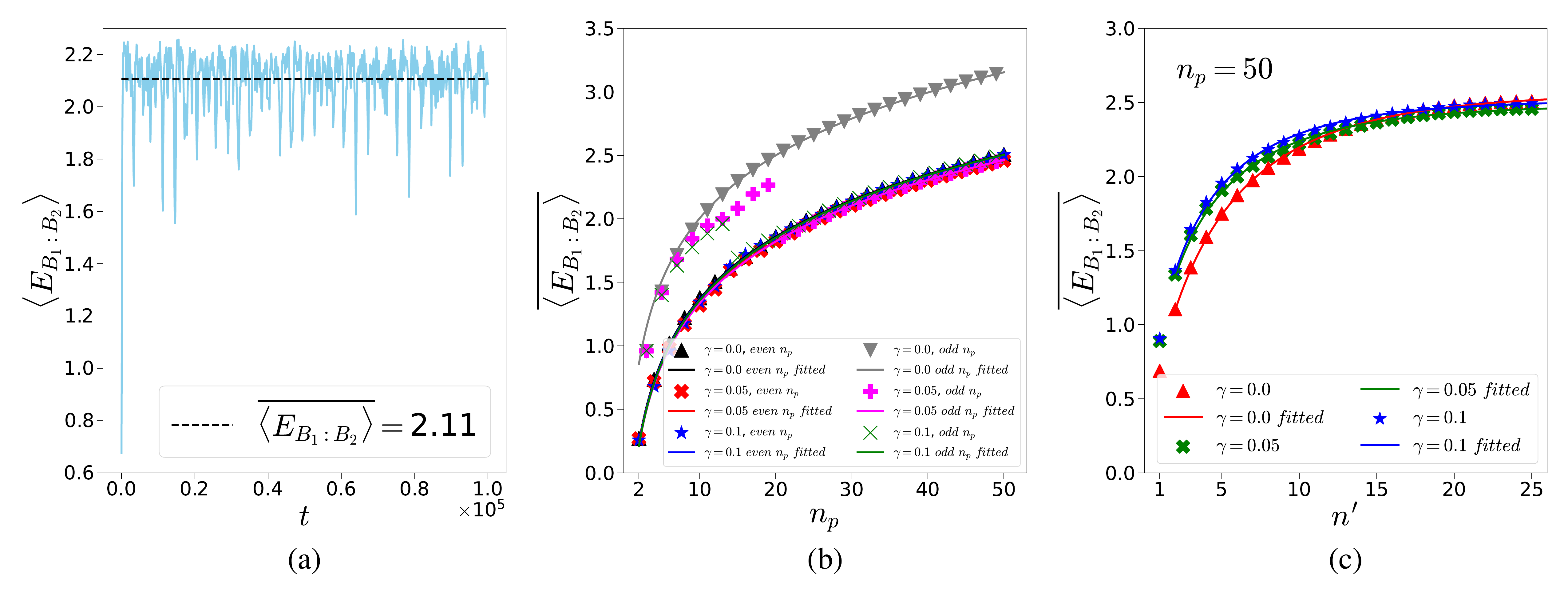}
    \caption{(a) A typical snapshot of the time-variation of $\langle E_{B_1:B_2}\rangle$. The horizontal line represent $\overline{\langle E_{B_1:B_2}\rangle}$. Variation of $\overline{\langle E_{B_1:B_2}\rangle}$ with (b) $n_p$ and (c) $n^{\prime}$ for different values of $\gamma$. For (b), the data is fitted to Eq.~(\ref{eq:entanglement_fit_n}) for $n_p\geq 24$  with fitting parameters ($a=-0.24(9),b=0.47(9)$) for $\gamma=0.05$ and even $n_p$, ($a=-0.27(6),b=0.485$) for $\gamma=0.05$ and odd $n_p$, ($a=-0.26(7),b=0.49(0)$) for $\gamma=0.1$ and even $n_p$ and ($a=-0.28(3),b=0.493$) for $\gamma=0.1$ and odd $n_p$. The data in (b) are fitted to Eq.~(\ref{eq:entanglement_fit_nprime}) with fitting paramters ($a=2.48(8),b=0.47(0),m=0.6(8)$) for $\gamma=0.05$ and ($a=2.51(7),b=0.47(3),m=0.7(0)$) for $\gamma=0.1$. The fitting parameters for $\gamma=0$ can be found in Fig.~\ref{fig:zero_field_dynamic}. All quantities plotted are dimensionless.}
    \label{fig:smalln0_dyn}
\end{figure*}

\paragraph{Variation with subsystem-size.} In the above scenario, $\langle E_{B_1:B_2}\rangle$ is varied effectively with the size of the full system. In contrast, we now consider a case where $\langle E_{B_1:B_2}\rangle$ is varied with increasing size of a subsystem of the full system, keeping the size of the full system fixed. For a fixed periphery-size $n_p$ (and subsequently for a fixed system-size $n$) and a varying partition size $n^\prime$ where the maximum value of $n^\prime$ is typically $n_p/2$,  for all $\gamma$,  $\langle E_{B_1:B_2}\rangle$ vary as 
\begin{eqnarray}
    \langle E_{B_1:B_2}\rangle &\sim& a\left[1-\exp\left(-b(n^\prime)^m\right)\right],
    \label{eq:entanglement_fit_nprime}
\end{eqnarray}
which is demonstrated in Fig.~\ref{fig:zero_field_dynamic}. Note that for  $\gamma=0$, the fitting parameters $a, b$ and $m$ remains unaffected up to three decimal places for $\langle E_{B_1:B_2}\rangle$ when $0\leq \Delta< 1$. 
On the other hand, for $\gamma \neq 0$, all the fitting parameters depends on $\Delta$.

\begin{figure*}
    \centering
    \includegraphics[width=0.7\textwidth]{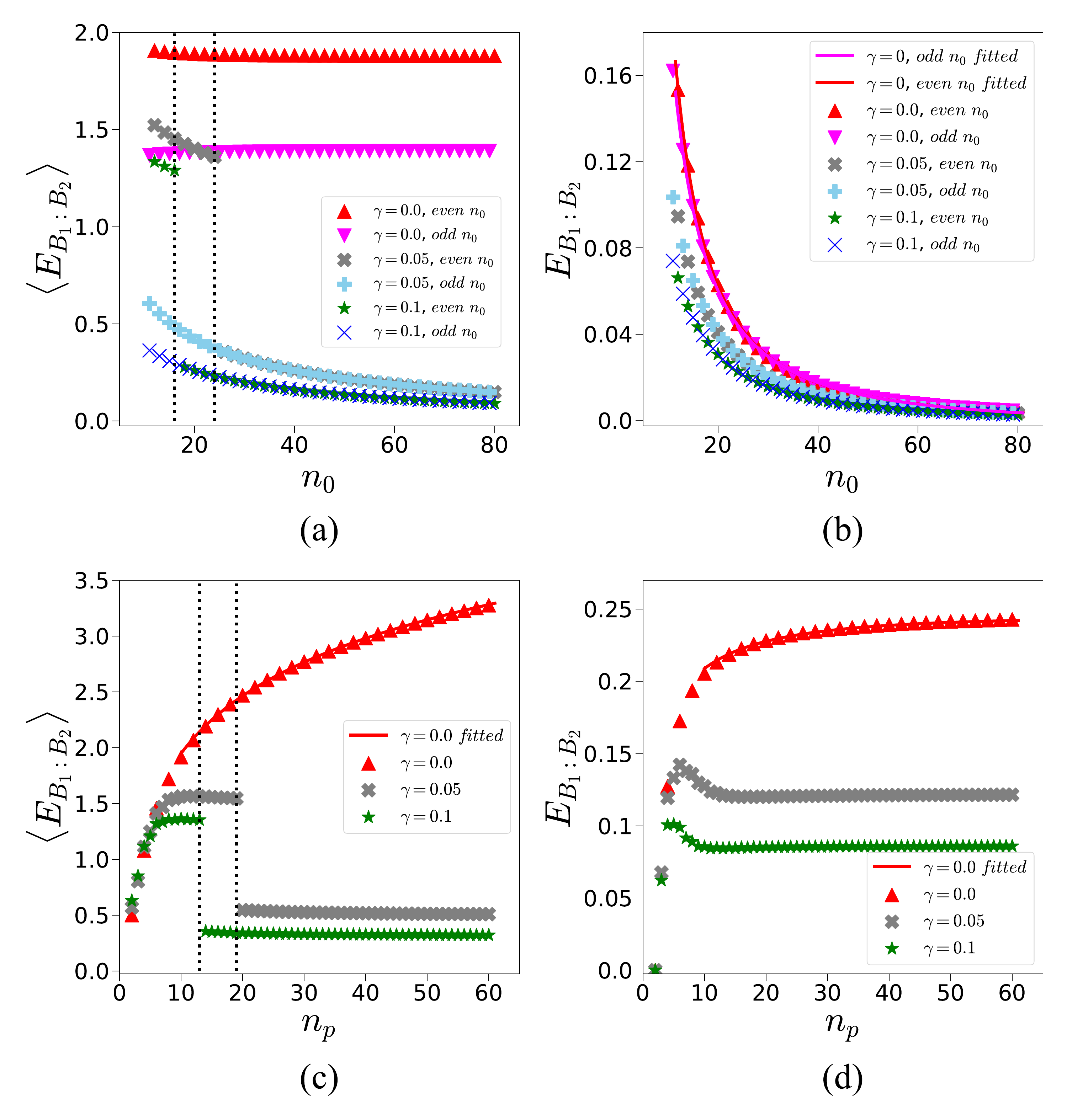}
    \caption{Variation of (a) $\langle E_{B_1:B_2}\rangle$ and (b) $E_{B_1:B_2}$ with $n_0$ in the limit $n_0/n_p \gg 1$, where $n_p$ is fixed at $n_p=10$. The onset of the degeneracy occurs for $n_0=24$ for $\gamma=0.05$ and $n_0=16$ for $\gamma=0.1$ in (a). Variation of (c) $\langle E_{B_1:B_2}\rangle$ and (d) $E_{B_1:B_2}$ with $n_p$ in the limit $n_0/n_p \rightarrow 1$, where $n_p=n_0$. The onset of the degeneracy occurs for $n_p=19$ for $\gamma=0.05$ and $n_p=13$ for $\gamma=0.1$ in (c). The data in (b), for $\gamma=0$ is fitted to Eq.~(\ref{eq:fit_ent_third_limit}) with the fitting parameters ($a=0, b=14(.2), m=1.8(1)$) for even $n_0$ and ($a=0, b=10(.6), m=1.7(3)$) for odd $n_0$. The $\gamma=0$ curve in (c) is fitted to Eq.~(\ref{eq:entanglement_fit_n}) with the fitting parameters ($a=0.23(5), b=0.51(5)$). The data in (d) is fitted to Eq.~(\ref{eq:entanglement_fit_np}) with fitting parameters ($a=0.244 ,b=0.53(2), m=0.54(1)$) for $\gamma=0$. All quantities plotted are dimensionless.}
    \label{fig:other_limits}
\end{figure*}

\noindent\textbf{Note 1: Effect of magnetic field.} We point out here that one can also introduce a magnetic field term in Eq.~(\ref{eq:star-ham}), so that the system Hamiltonian becomes $H_s+h\sum_{i=0}^{n_0+n_p-1}S_i^z$, where $h\rightarrow h/K$ is the strength of the magnetic field on each spin scaled by $|K|$, the energy scale of the system. In the limit $h\rightarrow\infty$, the system is fully polarized, and the state of the system is a product state with vanishing entanglement irrespective of the system size and anisotropy parameters, while the limit $h\rightarrow 0$ corresponds to the results discussed in this section. On the other hand, when $h\sim 1$, the finite-sized system undergoes different energy level crossings, and the ground state depends on the chosen value of $h$, while the values of $h$ at which these level crossings take place depend on $n_p$.  We, however, find that the overall \emph{qualitative} nature of the growth of LBPE with $n_p$ for both the cases of $\gamma=0$ and $\gamma\neq 0$ are similar to the respective cases reported for the limit $h\rightarrow 0$, although a more quantitative estimation through fitting of the data is difficult due to the discontinuous nature of the graphs owing to the energy level crossings and finite-size effects. 

\noindent\textbf{Note 2: Partial trace-based entanglement.} Note further that one can also determine bipartite entanglement on a bipartiton $B_1:B_2$ of the peripheral spins by computing an entanglement measure, $E$ (the same entanglement measure as used in Sec.~\ref{sec:static_entanglement}, see Eq.~(\ref{eq:avg_ent})), on the state $\rho_{B_1:B_2}=\text{Tr}_{A}\left[\rho_{AB}\right]$ as $E_{B_1:B_2}=E(\rho_{B_1:B_2})$. Our analysis indicates that the qualitative behaviours of $E_{B_1:B_2}$ with varying system sizes are the same as that of the LBPE $\langle E_{B_1:B_2}\rangle$, given by Eqs.~(\ref{eq:entanglement gap fitt}), (\ref{eq:entanglement_fit_n}), and (\ref{eq:entanglement_fit_nprime}).

\subsection{Field-induced dynamics in large periphery limit}
\label{subsec:field_induced_dynamics}

We now consider a situation where a time-dependent local magnetic field of strength $f(t)$ is applied to all spins in the star network. The field strength is such that 
\begin{eqnarray}
    f(t)=\left\{
    \begin{array}{cc}
    0; & t=0\\
    h; & t>0
    \end{array}
    \right.,
\end{eqnarray}
where $h>0$ (see~\cite{Barouch1970} for use of similar magnetic field). The system at $t>0$ therefore evolves in time due to the time-independent Hamiltonian
\begin{eqnarray} 
H_{\text{tot}}=H_s+h\sum_{i=0}^{n_0+n_p-1}S_i^z, 
\label{eq:total_Hamiltonian}
\end{eqnarray} 
where we redefine $h\rightarrow h/|K|$ to keep $H_{\text{tot}}$ dimensionless. Note that the features of block-diagonalization of $H$ discussed in Sec.~\ref{subsubsec:diagonalization} remain unchanged for $H_{\text{tot}}$ also,  irrespective of the value of $h$.

Starting from an initial state $\rho_{0}$ of the system, the time-evolved state of the full system at time $t$ is given by 
\begin{eqnarray}
    \rho(t)= \text{e}^{-\text{i}H_{\text{tot}}t}\rho_0\text{e}^{\text{i}H_{\text{tot}}t}.
\end{eqnarray}
The time-dependence of the LBPE can be computed by performing single-qubit projective measurements on the central qubit,  while the partial trace-based entanglement on the peripheral qubits can be computed using
$\rho_p(t)=\text{Tr}_{0}\rho(t)$, when the system is in state $\rho(t)$. In Fig.~\ref{fig:smalln0_dyn}(a), we plot the time-evolution of  $\langle E_{B_1:B_2}\rangle$ for a system with $n_p=30$, and $n^\prime=15$. The LBPE, $\langle E_{B_1:B_2}\rangle$, oscillates with $t$, and its average value, $\overline{\langle E_{B_1:B_2}\rangle}$, is obtained by averaging over the time-series data in the long-time limit. We probe the long-time limit of the time-evolving spin-star system using  $\overline{\langle E_{B_1:B_2}\rangle}$. Note that for $\gamma=0$, the field term commutes with the Hamiltonian irrespective of $\Delta$, leading to no time-evolution of the initial state. Therefore, we specifically focus on the cases with $\gamma\neq 0$. In Fig.~\ref{fig:smalln0_dyn}(b), we plot $\overline{\langle E_{B_1:B_2}\rangle}$ as function of $n_p$, where the partition-size $n^\prime=n_p/2$ $((n_p-1)/2)$ for even (odd) $n_p$. In contrast to the static case, for large $n_p$, $\overline{\langle E_{B_1:B_2}\rangle}$ exhibits logarithmic growth with $n_p$ in both $xy$-isotropic ($\gamma=0$) and $xy$-anisotropic ($\gamma\neq 0$) conditions, thereby negating the \emph{anisotropy effect}. The discrepancies in the logarithmic variation due to the finite-size of the system vanishes within $n_p\sim 20$. We further test the variations of  $\overline{\langle E_{B_1:B_2}\rangle}$ with $n^\prime$ for a fixed $n_p$, and found them to be similar to  the static case (see Eq.~(\ref{eq:entanglement_fit_nprime})) (see Fig.~\ref{fig:smalln0_dyn}(c)). Also, in alignment with the results from static case (\textbf{Note 2}, Sec.~\ref{subsec:static}), partial trace-based average entanglement,  $\overline{E_{B_1:B_2}}$, over the peripheral qubits exhibit qualitatively similar features as  $\overline{\langle E_{B_1:B_2}\rangle}$.

\begin{figure}
    \centering
    \includegraphics[width=.5\linewidth]{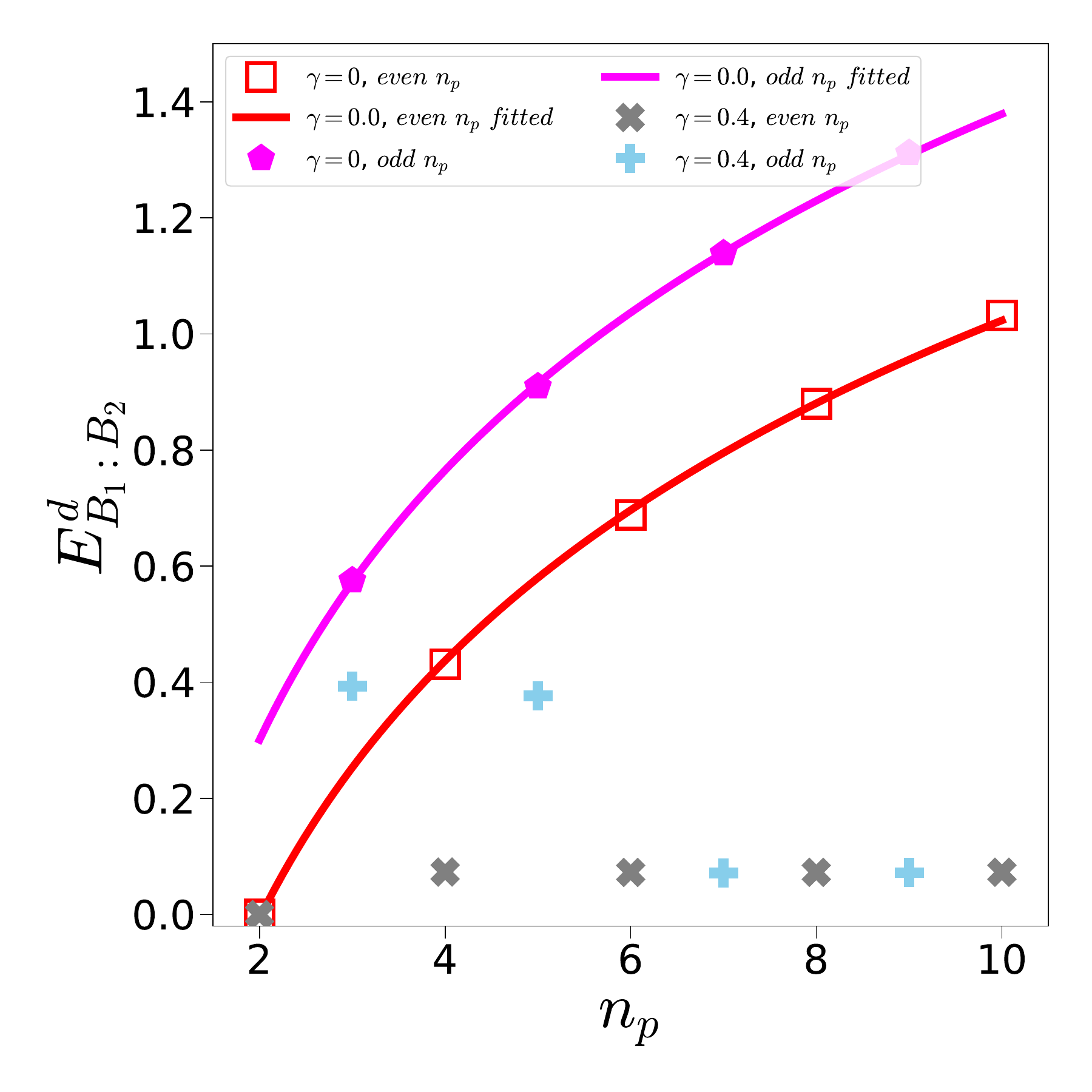}
    \caption{Variation of $E^{d}_{B_{1}:B_{2}}$ with increasing $n_p$, where disorder is applied on the spin-spin interaction strength with $\sigma_K=0.1$. The fitted lines are in accordance with Eq.~(\ref{eq:entanglement_fit_n}). The fitting parameters being $a=-0.4(4) (-0.1(6))$ $b=0.44(3) (0.46(4))$ for even (odd) $n_p$ and $\gamma=0$. All quantities plotted are dimensionless.}
    \label{fig:disorder}
\end{figure}

\subsection{Large and competing center limits}
\label{subsec:large_center}

A question that naturally arises at this point is whether the features of LBPE remain the same as one considers the limits $n_0/n_p\gg 1$, and $n_0/n_p\rightarrow 1$. In both cases, analysis with systems having small sizes ($n\leq 10$) indicates that the $S^z_i$ measurement on each qubit $i$ in the center is the optimal measurement for localizing maximum entanglement over any bipartition of the peripheral spins.  We first present the variation of $\langle E_{B_1:B_2}\rangle$ with $n_0$ for a fixed $n_p$ ($n_p=10$) in the limit $n_0/n_p\gg 1$ in Fig.~\ref{fig:other_limits}(a) (the minimal instance of two peripheral spins and $n_p$ central spins is represented in Fig.~\ref{fig:system}(b)). Our numerical analysis suggests that in contrast to the results obtained in the limit $n_0/n_p\ll 1$,
\begin{enumerate}
    \item for $\gamma\neq 0$, LBPE decreases monotonically and approaches zero asymptotically with increasing $n_0$ instead of exhibiting a saturation, and 
    \item for $\gamma=0$, LBPE exhibits a saturation as $n_0$ increases. 
\end{enumerate} 
Also, in contrast to the limit $n_0/n_p\ll 1$, in the present case, $E_{B_1:B_2}$ exhibits a different trend from that of $\langle E_{B_1:B_2}\rangle$, and decays monotonically with $n_0$ as (see Fig.~\ref{fig:other_limits}(b)) as
\begin{eqnarray}\label{eq:fit_ent_third_limit}
E_{B_1:B_2}\sim a+ bn_0^{-m},
\end{eqnarray}
where the fitting parameters are obtained from the numerical data. \JK{Note here that we have kept the parameter $a$ in the fitting function in order for addressing the possibility of $E_{B_1:B_2}$ decaying to a small yet non-zero positive value. However, $a$ is found to be zero in our numerical analysis.}

In the limit $n_0/n_p\rightarrow 1$, the structure of the ground states  with permutation symmetry in both central and peripheral qubits is given by 
\begin{eqnarray}
    \ket{\Psi_0}&=&\sum_{l^\prime=0}^{n_0}\sum_{l=0}^{n_p} c_{l,l^\prime}\ket{D_{l^\prime}^{n_0}}\otimes\ket{D^{n_p}_l},
    \label{eq:ground_state_form_coupled_central}
\end{eqnarray}
where $\sum_{l^\prime}\sum_l |c_{l,l^\prime}|^2=1$, and the definition of $\ket{D_{l^\prime}^{n_0}}$ is similar to that of $\ket{D_l^{n_p}}$. Similar to the case of single central qubit,  the TGS $\rho_0$ corresponding to the degenerate and the effectively degenerate ground states $\ket{\Psi^\pm_0}$ is of the form (\ref{eq:thermal_ground_state}), while the mixed state $\rho_p$ on the peripheral qubits obtained via tracing out the central qubits is of the form (\ref{eq:mixed_state_dicke_basis}). In Fig.~\ref{fig:other_limits}(c), we plot LBPE as a function of $n_p$ with $n_0=n_p$. For $\gamma\neq 0$ as well as for $\gamma=0$, the trends of LBPE is similar to the case of $n_0/n_p\ll 1$. In Fig.~\ref{fig:other_limits}(d), we plot the variations of $E_{B_1:B_2}$ as a function of $n_p$, keeping $n_0=n_p$. It is noteworthy that in contrast to the limit $n_0/n_p\ll 1$, $E_{B_1:B_2}$ for $\gamma=0$ saturates with increasing $n_p$ as 
\begin{eqnarray}
    E_{B_1:B_2} &\sim& a\left[1-\exp\left(-bn_p^m\right)\right].
    \label{eq:entanglement_fit_np}
\end{eqnarray}
For $\gamma\neq 0$ also, $E_{B_1:B_2}$ saturates with $n_p$, although the approach to saturation is different from that of the $\gamma=0$, and the saturation value decreases with increasing $\gamma$, implying an \emph{anisotropy effect}. Further, the behaviour of $E_{B_1:B_2}$ with $n_p$ is different from the same for $\langle E_{B_1:B_2}\rangle$, which is also in contrast to the results obtained in the limit $n_0/n_p\ll 1$.  

\noindent\textbf{Note 3: Field-induced dynamics.} One can also compute the long time-averaged localizable entanglement and partial trace-based entanglement in the limits $n_0/n_p\gg 1$ and $n_0/n_p\rightarrow 1$. Our analysis indicates that in the latter limit, turning on the magnetic field does not negate the anisotropy effect in the case of  the time-averaged partial trace-based bipartite peripheral entanglement, $\overline{E_{B_1:B_2}}$, while in the case of $\overline{\langle E_{B_1:B_2}\rangle}$, it does. This exhibits yet another example where the partial trace-based entanglement behaves differently than the localizable entanglement.

\begin{figure}
    \centering
    \includegraphics[width=0.45\linewidth]{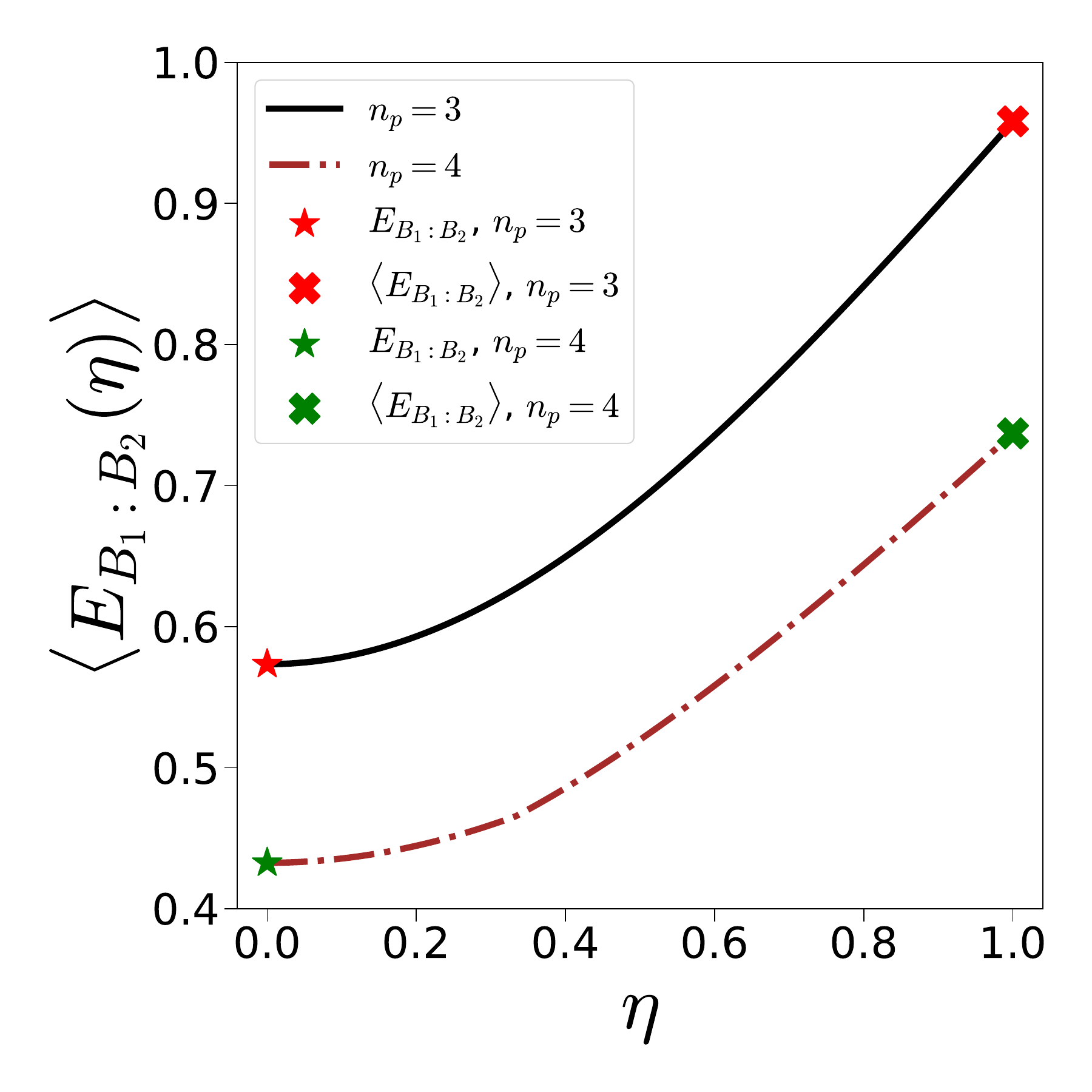}
    \caption{Variations of $\langle E_{B_1:B_2} (\eta)\rangle$ as a function of $\eta$ in the case of  spin-star systems with $n_0=1$ and $n_p=3$ ($n=4$), and $n_p=4$ ($n=5$). For $\eta=0$, $\langle E_{B_1:B_2}(\eta=0)\rangle=E_{B_1:B_2}$ -- the entanglement obtained from the partial trace-based approach, while for $\eta=1$, $\langle E_{B_{1}:B_2}(\eta=1)\rangle=\langle E_{B_1:B_2}\rangle$, the LE corresponding to the noiseless projection measurement on the central qubit. All quantities plotted are dimensionless. }
    \label{fig:bounds}
\end{figure}

\JK{
\subsection{Effect of disorder}
\label{subsec:disorder}

So far, we have considered the spin-spin interaction strength $K$ corresponding to all central-peripheral spin-pairs to be identical, and have chosen it to be $1$. However, in a realistic scenario, the interaction strength can be non-uniform, or \emph{disordered}~\cite{de_dominicis_giardina_2006}, and the values of $K$ can be chosen from a suitable distribution with a fixed mean. We consider a Gaussian distribution, $P(K)$, of fixed mean, $\langle K\rangle=1$ (the chosen value is motivated from the case of uniform interaction strengths in order for facilitating comparison), and fixed standard deviation, $\sigma_K$,  where $\sigma_K$ quantifies the \emph{strength} of the disorder. The collection of $n_p$ random values of $K$, one for each central-peripheral spin-pair $(0,j)$, $j=1,2,\cdots,n_p$ with $n_0=1$ represents a \emph{random parameter configuration} of the spin-star system. Note here that for a random realization $\{K\}$, the spin-spin interaction strengths corresponding to each central-peripheral spin-pair is different, implying $[H_s,\mathbf{J}_p^2]\neq 0$ and $[H_s,\mathbf{S}_0^2]\neq 0$, making diagonalization of the system difficult, and the methodology discussed in Sec.~II.A can not be applied. However, we tackle this problem numerically, and restrict ourselves to moderate system sizes such that $n_0=1$, and $n_p\leq 10$. Note further that due to the breaking of permutation symmetry, exact calculation of peripheral entanglement in a measurement-based approach also becomes computationally intractable even for moderate-sized systems due to the optimization involved in the definition, and the requirement of performing this optimization repeatedly for each configuration $\{K\}$ in a statistically large number of configurations $\{K\}$. However, with the available numerical resource, for moderate system sizes, we compute the partial trace-based peripheral entanglement, eg.,  $E_{B_1:B_2}(\{K\})$ for each such random realization $\{K\}$ of the system, and subsequently perform a \emph{quenched} average of $E_{B_1:B_2}(\{K\})$ over a statistically large number of random parameter setups. The quenched averaged entanglement is given by 
\begin{eqnarray}
E_{B_1:B_2}^d &=& \int E_{B_1:B_2}(\{K\}) P(\{K\})d\{K\},
\end{eqnarray}
where the superscript $d$ represents a quenched average, and $P((\{K\}))$ is the probability of occurrence of the random realization $\{K\}$. We present the data obtained through our numerical simulation in Fig.~\ref{fig:disorder}.  Note that a logarithmic growth with $n_p$ is observed even for $E_{B_1:B_2}^d$ in the case of moderate system sizes ($n_p\leq 10$, $n_0=1$), indicating a robustness of our result on the dependence of $E_{B_1:B_2}$ on $n_p$ for $n_0=1$ (the corresponding \emph{ordered} result, reported in Sec.~\ref{subsec:static}, can be obtained as a special case at $\sigma_K=0$).}

\section{Localization via unsharp measurements}
\label{sec:noisy_measurements}

So far, we have assumed the single-qubit measurements considered in Sec.~\ref{sec:static_entanglement} to be \emph{perfect}. However, in reality, measurement devices can be noisy, and a logical question is whether the trends of LE reported in Sec.~\ref{sec:static_entanglement} persist in situations where one is forced to apply \emph{noisy} optimal projection measurements on the central qubits.  
To test this, we consider the optimal measurement corresponding to the case of $\eta=1$ (perfect projective measurements), i.e., $S^z$ measurements on all central qubits, and allow values of $\eta$ in the range $0\leq\eta<1$ such that the measurements are noisy. In what follows, we discuss the trends of $\langle E_{B_1:B_2}(\eta)\rangle$ (see Eq.~(\ref{eq:LE}), Sec.~\ref{sec:static_entanglement}) against the system size.

\begin{figure*}
    \centering
    \includegraphics[width=0.8\linewidth]{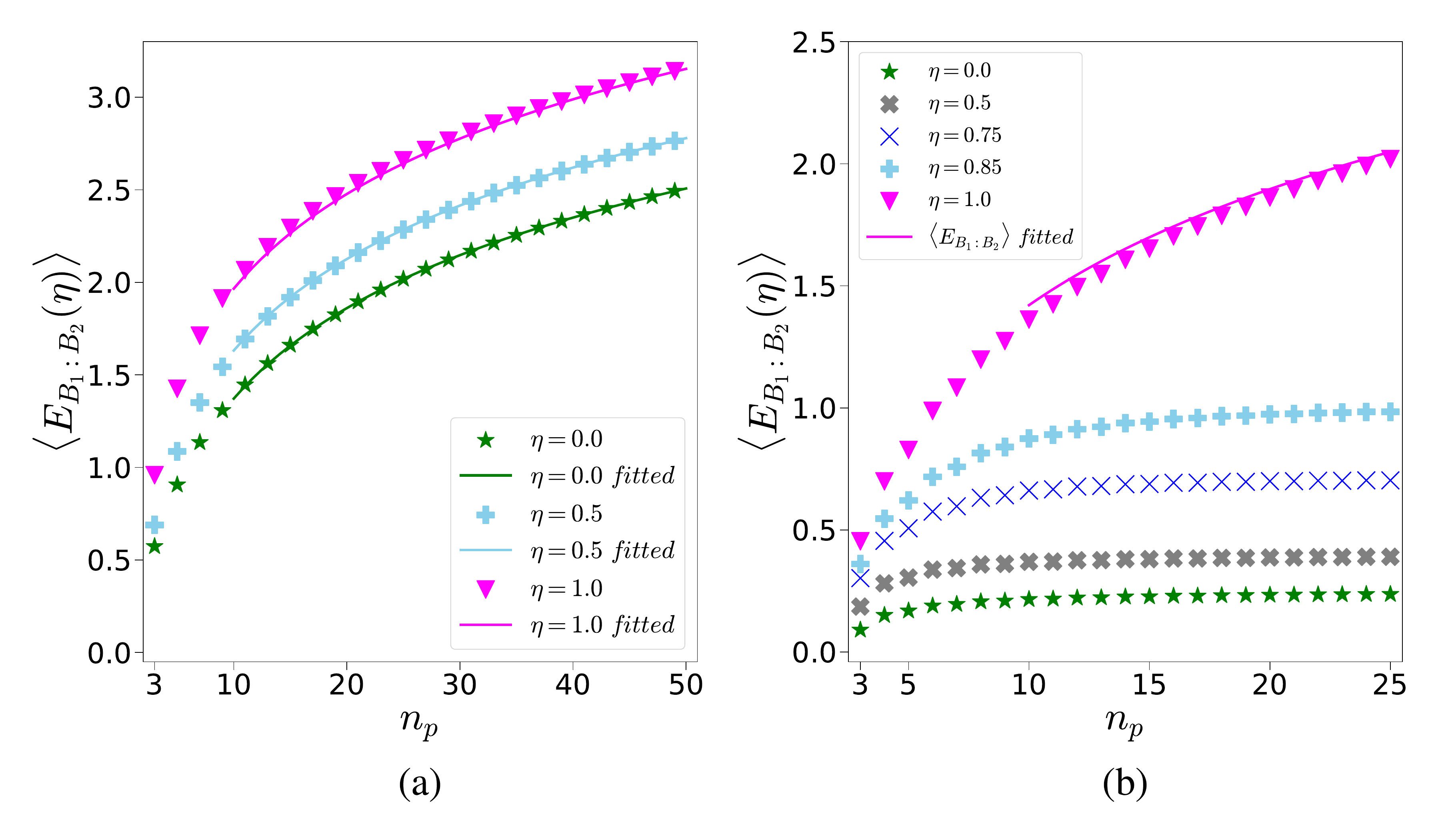}
    \caption{Variations of $\langle E_{B_1:B_2}(\eta)\rangle$ as a function of $n_p$ for different values of $\eta$ in the range $0\leq \eta\leq 1$ for the limits (a) $n_0/n_p\ll 1$ and (b) $n_0/n_p\rightarrow 1$ of the spin-star system. In (a), the data for $\langle E_{B_1:B_2}(\eta)\rangle$ is fitted to Eq.~(\ref{eq:entanglement_fit_n}), where only the case of odd $n_p$ is shown, with the fitting parameters corresponding to $\eta=1$ available in Fig.~\ref{fig:zero_field_dynamic}(a), and the same corresponding to $\eta=0$ ($\eta=0.5$) are $a=-0.25(3),b=0.489$ ($a=-0.01,b=0.494$) . In (b), only the data corresponding to $\langle E_{B_1:B_2}(\eta=1)\rangle$ are fitted to Eq.~(\ref{eq:entanglement_fit_n}), and the fitting parameters are given by $a=-0.15(7),b=0.47(5)$. All quantities plotted are  dimensionless.}
    \label{fig:log_noise}
\end{figure*}

We first take up the large periphery limit, and fix $n_0=1$ to write the ground state (Eq.~(\ref{eq:ground_state_form})) as 
\begin{eqnarray}\label{eq: state}
    \rho &=& \ket{0}\bra{0}\otimes \sum_{l,m}c_l c_m^* \ket{D^{n_p}_l}\bra{D^{n_p}_m}+\ket{1}\bra{1}\otimes \sum_{l^{\prime},m^{\prime}}d_{l^{\prime}} d^*_{m^{\prime}} \ket{D^{n_p}_{l^{\prime}}}\bra{D^{n_p}_{m^{\prime}}}+\ket{0}\bra{1}\otimes \sum_{l,l^{\prime}}c_{l} d^*_{l^{\prime}} \ket{D^{n_p}_l}\bra{D^{n_p}_{l^{\prime}}}\nonumber\\&&+\ket{1}\bra{0}\otimes \sum_{l,l^{\prime}} d_{l^{\prime}}c_{l}^* \ket{D^{n_p}_{l^{\prime}}}\bra{D^{n_p}_{l}},
\end{eqnarray}
where we have used $\ket{\psi_0}=\ket{0}$ and $\ket{\psi_0^\perp}=\ket{1}$ since $S^z$ measurement is optimal in the case of $\eta=1$. Further, $M(\eta,\lambda_0,\hat{k}_0)$  can be written as 
\begin{eqnarray}
    M_\pm(\eta) &=& \sqrt{\frac{1\pm\eta}{2}}\ket{0}\bra{0} + \sqrt{\frac{1\mp\eta}{2}}\ket{1}\bra{1},
\end{eqnarray}
where we write $M(\eta,\pm 1,\hat{k}_0)$ as $M_\pm(\eta)$, using the fact that $\hat{k}_0$ corresponds to $S^z_0$ measurement on the central qubit. Using $M_\pm(\eta)$, the post-measured states on the spin-star system corresponding to the measurement outcomes $\pm 1$, are given, up to normalization, by 
\begin{eqnarray}
        \varrho^{(\pm)} &=& \frac{1\pm\eta}{2}\ket{0}\bra{0}\otimes \sum_{l,m}c_l c_m^* \ket{D^{n_p}_l}\bra{D^{n_p}_m}+\frac{1\mp\eta}{2}\ket{1}\bra{1}\otimes \sum_{l^{\prime},m^{\prime}}d_{l^{\prime}} d^*_{m^{\prime}} \ket{D^{n_p}_{l^{\prime}}}\bra{D^{n_p}_{m^{\prime}}}\nonumber \\&&+\sqrt{\frac{1-\eta^2}{4}}\Big[\ket{0}\bra{1}\otimes \sum_{l,l^{\prime}}c_{l} d^*_{l^{\prime}} \ket{D^{n_p}_l}\bra{D^{n_p}_{l^{\prime}}}+\ket{1}\bra{0}\otimes \sum_{l,l^{\prime}} d_{l^{\prime}}c_{l}^* \ket{D^{n_p}_{l^{\prime}}}\bra{D^{n_p}_{l}}\Big],
\end{eqnarray}
which occurs with the probability
\begin{eqnarray}
    p^{\pm}=\frac{1\pm\eta \alpha}{2},
\end{eqnarray}
with $\alpha=\sum_l |c_l|^2 - \sum_{l^{\prime}} |d_l^{\prime}|^2$. Tracing out the central qubit from $\varrho^{\pm}$ provides with the peripheral post-measured states $\varrho^{\pm}_p$ as 
\begin{eqnarray}\label{eq:post-mmt state}
        \varrho_p^{(\pm)} &=& \frac{1\pm\eta}{1\pm\eta \alpha}\sum_{l,m}c_l c_m^* \ket{D^{n_p}_l}\bra{D^{n_p}_m}+\frac{1\mp\eta}{1\pm\eta \alpha} \sum_{l^{\prime},m^{\prime}}d_{l^{\prime}} d^*_{m^{\prime}} \ket{D^{n_p}_{l^{\prime}}}\bra{D^{n_p}_{m^{\prime}}},  
\end{eqnarray}
which one can write as 
\begin{eqnarray}
        \varrho_p^{(\pm)} &=& \frac{(1\pm\eta)(1+\alpha)}{2(1\pm\eta \alpha)}  \tilde{\varrho}_p^{(+)} +\frac{(1\mp\eta)(1-\alpha)}{2(1\pm\eta\alpha)}\tilde{\varrho}_p^{(-)}
\end{eqnarray}
by recognizing $\tilde{\varrho}_p^{\pm}$ as $\varrho_p^{\pm}$ in Eq.~(\ref{eq:post-mmt state}) for $\eta=1$. Therefore, the entanglement localizable on the periphery via noisy $S^z$ measurement on the central qubit with a noise strength $\eta (\neq 1)$ is computed as 
\begin{eqnarray}
    \langle E_{B_1:B_2}(\eta)\rangle &=& \frac{1+\eta\alpha}{2} E_{B_1:B_2}^{(+)}+\frac{1-\eta\alpha}{2}  E_{B_1:B_2}^{(-)},
\end{eqnarray}
where we write $E(\varrho_p^{(\pm)})$ as $E_{B_1:B_2}^\pm$ to remind ourselves that the bipartite entanglement is computed over the bipartition $B_1:B_2$ on the periphery. Observe that 
\begin{eqnarray}
\langle E_{B_1:B_2}(\eta=1)\rangle = &=& \frac{1+\alpha}{2} E_{B_1:B_2}^{(+)}+\frac{1-\alpha}{2}  E_{B_1:B_2}^{(-)}
\label{eq:LE_perfect}
\end{eqnarray}
provides the localizable entanglement corresponding to the perfect projection measurement on the central qubit.

\noindent\textbf{Note 4: Optimization of LE.} We point out here that the above calculation is performed in view of the situation where one knows $S^z$ measurement to be optimal on the central qubit, and is yet forced to perform a noisy $S^z$ measurement due to a faulty measurement apparatus. One can also perform an optimization of the average entanglement over all directions of the unsharp measurements, when the noise strength is fixed at $\eta$. In the present case, the optimization is found to occur corresponding to $S^z$ measurement for all values of $\eta$ in the range $0< \eta<1$.  

\noindent\textbf{Note 5: Using a convex entanglement measure.} One can further consider a \emph{convex} measure $E$ of entanglement~\cite{horodecki2009}, resulting in (see Eq.~(\ref{eq:LE_perfect}))
\begin{eqnarray}
    \langle E_{B_1:B_2}(\eta)\rangle &\leq& \frac{1+\alpha}{2}E_{B_1:B_2}^{(+)}+\frac{1-\alpha}{2}E_{B_1:B_2}^{(-)} =\langle E_{B_1:B_2}(\eta=1)\rangle,
    \label{eq:upper_bound}
\end{eqnarray}
implying that $\langle E_{B_1:B_2}(\eta)\rangle$ is bounded above by the localizable entanglement in the noiseless scenario, given by $\langle E_{B_1:B_2}(\eta=1)\rangle$. Further, using the convexity of $E$, one can obtain a lower bound of $\langle E_{B_1:B_2}(\eta)\rangle$ as   
\begin{eqnarray}
    \langle E_{B_1:B_2}(\eta)\rangle &\geq& E\left(\frac{1+\alpha}{2} \tilde{\varrho}_p^{(+)} + \frac{1-\alpha}{2} \tilde{\varrho}_p^{(-)}\right)
    =\langle E_{B_1:B_2}(\eta=0)\rangle =E(\rho_p),
    \label{eq:lower_bound}
\end{eqnarray}
where we recognize 
\begin{eqnarray}
    \rho_p=\text{Tr}_0\rho=\frac{1+\alpha}{2} \tilde{\varrho}_p^{(+)} + \frac{1-\alpha}{2} \tilde{\varrho}_p^{(-)}
\end{eqnarray}
due to the basis-independence of partial trace.
 
Note that the logarithmic negativity, used as the entanglement measure of choice so far, is \emph{not} a convex measure of entanglement~\cite{plenio2005}. However, our numerical analysis suggests that the upper and the lower bounds of $\langle E_{B_1:B_2}(\eta)\rangle$, for all values of $\eta$ in the range $0<\eta<1 $ remains valid even when logarithmic negativity is used as the entanglement measure (see Fig.~\ref{fig:bounds}). Further, for all values of  $\eta$ in the range $0<\eta<1 $, $\langle E_{B_1:B_2}(\eta)\rangle$ follows a logarithmic trend with increasing periphery-size, $n_p$, as reported in the noiseless scenario discussed in Sec.~\ref{subsec:static}. See Fig.~\ref{fig:log_noise}(a) for a demonstration.

\textbf{Note 6: Large and competing center limits.} The calculation for the case of the single central spin, as demonstrated above, can be generalized for the cases of $n_0>1$, and the bounds reported in Eqs.~(\ref{eq:lower_bound}) and (\ref{eq:upper_bound}) hold for the limits $n_0/n_p\gg 1$ and $n_0/n_p\rightarrow 1$ also. However, in contrast to the the limit $n_0/n_p\ll 1$, when $n_0\approx n_p$, $\langle E_{B_1:B_2}\rangle$ does not follow a logarithmic trend with increasing $n_p$   (see Sec.~\ref{subsec:large_center}). In Fig.~\ref{fig:log_noise}(b), we plot $\langle E_{B_1:B_2}(\eta) \rangle$ as a function of $n_p$ for different values of $\eta$ in the range $0\leq \eta\leq 1$, where a transition of the variation of  $\langle E_{B_1:B_2}(\eta) \rangle$ with $n_p$ from a non-logarithmic to a logarithmic trend with increasing $\eta$ is apparent.

\section{Conclusion and Outlook}
\label{sec:conclude}

\JK{In this paper, we study the XYZ model on a spin-star system of size $n (=n_0+n_p)$, where $n_0$ central spins are connected to $n_p$ peripheral spins in such a way that each central spin interacts with all peripheral spins. We consider the small-, large-, and competing-center limits of the system described respectively by $n_0/n_p\ll1$, $n_0/n_p\gg1$, and $n_0/n_p\rightarrow1$. We show that the ground state of the system is doubly degenerate when the system-size is odd, while there is a finite energy gap between the ground and the first excited states of the system with even number of spins. This energy gap decreases rapidly with increasing system-size, and finally vanishes, inducing an \emph{effective} double degeneracy of the ground states. In the limit $n_0/n_p\ll 1$, our numerical analysis demonstrates two different approach of the energy gap to zero with increasing periphery size in the case of the $xy$-isotropic ($\gamma=0$) and anisotropic ($\gamma\neq 0$) interactions between spin pairs belonging to the center and the periphery. In contrast, in the limit $n_0/n_p\rightarrow 1$, the energy gap in the $\gamma=0$ case saturates to a non-zero value, while for $\gamma\neq 0$, the energy gap vanishes asymptotically with increasing periphery size, following the trend of isotropic case in the limit $n_0/n_p\ll 1$.       

In the limit $n_0/n_p\ll 1$, we compute localizable bipartite entanglement over typically equal bipartition of the periphery, which exhibits a logarithmic growth with the periphery-size when the system is $xy$-isotropic (i.e., $\gamma=0$), but for $\gamma\neq 0$, it does not. We call this to be the \emph{anisotropy effect}, which is shown to be nullified by kick-starting a magnetic field of constant strength on all spins in the system, where the average values of the time-series of localizable bipartite peripheral entanglement exhibits a logarithmic growth with the periphery-size irrespective of the values of $\gamma$. We further consider the large center ($n_0/n_p\gg 1$) and the competing center ($n_0/n_p\rightarrow 1$) limits of the model, and demonstrate that for a fixed periphery and an increasing center size, the localizable bipartite peripheral entanglement asymptotically vanishes (saturates to a constant value) for $\gamma\neq 0$ ($\gamma=0$). 

Besides localizable bipartite peripheral entanglement, we check the bipartite entanglement on the equal bipartition of the periphery computed using a partial trace-based approach, and find its trend with $n_p$ to be the same with the localizable bipartite peripheral entanglement in the $n_0/n_p\ll 1$ limit of the model, but not in the other two limits. For moderate-sized systems, the results for the bipartite peripheral entanglement computed using the partial trace-based method are found to be immune to disorder in the spin-spin interactions strengths. Further, assuming white noise in the measurement apparatus and with an appropriately generalized definition of the localizable entanglement, we show that in the limit $n_0/n_p\ll 1$, localizable bipartite peripheral entanglement corresponding to all possible noise strength exhibits a logarithmic growth with the periphery-size as in the case of the noiseless scenario. However, in contrast, in the limit $n_0/n_p\rightarrow 1$, it does not.}    

\JK{A number of possible research directions open up from this study. Within the range of the results reported in this paper, it would be interesting to investigate whether similar features in entanglement exists if interactions are present within the central and/or the peripheral spins themselves~\cite{Richter_1994}. The challenge lies in the computation of entanglement, as the states on the peripheral spins would not belong to the subspaces of the Dicke states. Also, a similar investigation  in a star-chain configuration of spin-$\frac{1}{2}$ particles~\cite{Yao2011} is yet to be performed. Besides, similarity of our results with the already existing results coresponding to the entanglement computed using partial trace-based approach for the LMG model~\cite{Latorre2005} highlights the question as to what would be the underlying principle to model a spin-lattice system that exhibits similar entanglement properties. Also, the implications of the reported entanglement properties of the spin-star model with respect to the use of the model in quantum protocols such as switch in quantum networks~\cite{Yung_2011} and quantum state transfer~\cite{Deng_2008} are yet to be explored. Moreover, recent interests in exploring quantum technological aspects involving $d$-level systems, or qudits~\cite{Wang2020} highlights the relevance of exploring similar models where lattice sites are populated with spins having a higher spin quantum number.}

\acknowledgments 

The Authors acknowledge the use of \href{https://github.com/titaschanda/QIClib}{QIClib} -- a modern C++ library for general purpose quantum information processing and quantum computing. HKJ thanks the Prime Minister Research Fellowship (PMRF) program, Government
of India, for the financial support. AKP
acknowledges the support from the Anusandhan National Research Foundation (ANRF) of the Department of Science and Technology (DST), India, through the Core Research Grant (CRG) (File No. CRG/2023/001217, Sanction Date 16 May 2024). The Authors also thank Aditi Sen (De) for useful discussions.

\onecolumngrid 

\appendix 

\section{Small systems with single central spin}
\label{app:small_system}

In this section, we look into two specific examples for $n_p=2$ and $n_p=3$ in the case of a star network of $(n_p+1)$ spins, as follows.

\subsection{\texorpdfstring{$n_p=2$}{np=2}}

We first consider the case of $J_{p}=1$  ($m_p=-1,0,1$) among the allowed values $J_p=0,1$ for $n_p=2$. We group the basis states according to Eq.~(\ref{eq:grouping}) as $    \{\ket{-1/2}\ket{-1}, \ket{1/2}\ket{0} ,\ket{-1/2}\ket{1} \}$ and $\{\ket{1/2}\ket{-1}, \ket{-1/2}\ket{0} ,\ket{1/2}\ket{1}\}$ respectively, leading to the $\mathcal{H}_{1}$ block having the form (\ref{eq:block_diagonal_forms}),  with $A_1$ and $A_2$ being two $3\times 3$ matrices given by 
\begin{eqnarray}
A_1 &=& \frac{1}{2} \begin{bmatrix}
    \Delta  &  \gamma \alpha^{+}_{0}   &   0  \\
    \gamma \alpha^{-}_{1}  &  0 &  \alpha^{+}_{-1}   \\
    0    &    \alpha^{-}_{0}    &    -\Delta 
\end{bmatrix}, 
A_2 = \frac{1}{2}
\begin{bmatrix}
   -\Delta  &  \alpha^{+}_{0}   &  0   \\
 \alpha^{-}_{1}  &  0 &  \gamma \alpha^{+}_{-1}    \\
    0    &    \gamma \alpha^{-}_{0}     &    \Delta 
\end{bmatrix}, 
\end{eqnarray} 
upto a normalization factor, where 
\begin{eqnarray}
\alpha^{\pm}_{m_{p}} = \sqrt{J_{p}(J_{p}+1) - m_{p}(m_{p}\pm 1)}.
\label{eq:alpha}
\end{eqnarray}
Note that $|A_1|=|A_2|=2\Delta(\gamma^2-1)$, which vanishes for $\Delta=0$ or $\gamma=\pm 1$, implying that at least one of the eigenvalues of both $A_1$ and $A_2$ vanishes at these conditions. For $\Delta=0$, the eigenvalues are $0,\pm\sqrt{2(\gamma^{2}+1)}/2$ for both  $A_{1}$ and $A_{2}$, leading to the doubly-degenerate ground state energy $-\sqrt{2(\gamma^2+1)}/2$. The corresponding eigenvectors belonging to the blocks $A_1$ and $A_2$ are given by
\begin{eqnarray}
    \ket{\psi}_{0,{A_{1}}} &=& \frac{1}{\sqrt{2(\gamma^{2}+1)}} \ket{-1/2}_0\Big(\gamma\ket{-1}_p+\ket{1}_p\Big)-\frac{1}{\sqrt{2}}\ket{1/2}_0\ket{0}_p,\nonumber\\
    \ket{\psi}_{0,{A_{2}}} &=& \frac{1}{\sqrt{2(\gamma^{2}+1)}} \ket{1/2}_0\Big(\ket{-1}_p+\gamma\ket{1}_p\Big)-\frac{1}{\sqrt{2}}\ket{-1/2}_0\ket{0}_p,
\end{eqnarray} 
where clearly $\ket{\psi}_{0,A_1}=\mathcal{O}\ket{\psi}_{0,A_2}$, and vice-versa (see Sec.~\ref{subsubsec:single_central_spin_ground_state}). Consequently, the doubly degenerate ground-states corresponding to the ground-state energy $-\sqrt{2(\gamma^2+1)}/2$ and in the common eigenspace of $H_s$ and $\mathcal{O}$ are  given by Eq.~(\ref{eq:degenerate_ground_states}).   Similarly for  $\gamma=\pm1$, the eigenvalues of either of $A_1$ and $A_2$ are $0,\pm\sqrt{4+\Delta^{2}}/2$, with the doubly-degenerate ground state energy $-\sqrt{4+\Delta^2}/2$, and the corresponding eigenvectors being  
\small 
\begin{eqnarray}
    \ket{\psi}_{0,{A_{1}}} &=& \Big(\frac{\sqrt{2}}{\Delta-\sqrt{4+\Delta^{2}}}\Big) \ket{-1/2}_{0}\ket{1}_{p}  -\Big(\frac{\sqrt{2}}{\Delta+\sqrt{4+\Delta^{2}}}\Big)\ket{-1/2}_{0}\ket{-1}_{p}+  \ket{1/2}_{0}\ket{0}_{p},\nonumber\\
   \ket{\psi}_{0,{A_{2}}} &=& \Big(\frac{\sqrt{2}}{\Delta-\sqrt{4+\Delta^{2}}}\Big) \ket{1/2}_{0}\ket{-1}_{p}    -\Big(\frac{\sqrt{2}}{\Delta+\sqrt{4+\Delta^{2}}}\Big)\ket{1/2}_{0}\ket{1}_{p}+ \ket{-1/2}_{0}\ket{0}_{p},\nonumber\\ 
\end{eqnarray} \normalsize 
up to a normalization factor. It is easy to see that similar to the case of $\Delta=0$, $\ket{\psi}_{0,A_1}$ and $\ket{\psi}_{0,A_2}$ are connected via $\mathcal{O}$, and the doubly-degenerate ground states in the common eigenspace of $H_s$ and $\mathcal{O}$ are given by Eq.~(\ref{eq:degenerate_ground_states}), as in the case of $\Delta=0$. The double degeneracy in ground state is maintained in the entire range $0\leq |\gamma|\leq 1$, $0\leq |\Delta|\leq 1$.  

We point out here that all sectors of $H_s$ corresponding to values of $J_p\neq n_p/2$ can be block-diagonalized using a similar grouping of basis elements.  In the present example, the block $\mathcal{H}_0$ corresponding to $J_p=0$, is diagonal in the basis  $\{\ket{\frac{1}{2}}\otimes\ket{0}, \ket{-\frac{1}{2}}\otimes \ket{0}\}$, with both eigenvalues vanishing irrespective of the values of $(\gamma,\Delta)$. 

\subsection{\texorpdfstring{$n_p=3$}{np=3}}

In this case, $J_p=3/2,1/2,1/2$, and we start with the block $H_s(J_p=3/2)$, where grouping the basis as $\{ \ket{-1/2}\ket{3/2}, \ket{1/2}\ket{1/2}, \ket{-1/2}\ket{-1/2},  \ket{1/2}\ket{-3/2}\}$ and $\{\ket{1/2}\ket{3/2}, \ket{-1/2}\ket{1/2}, \ket{1/2}\ket{-1/2},  \ket{-1/2}\ket{-3/2}\}$ makes $\mathcal{H}_{3/2}$ of the form~(\ref{eq:block_diagonal_forms}) with the $4\times4$ matrices $A_1$ and $A_2$ given by 
\begin{eqnarray}
A_1&=& \frac{1}{2}
\begin{bmatrix}
    -\frac{3\Delta}{2}  &  \alpha^{-}_{\frac{3}{2}}   &   0   &    0   \\
    \alpha^{+}_{\frac{1}{2}}  &  \frac{\Delta}{2}   &  \gamma\alpha^{+}_{-\frac{1}{2}} & 0   \\
    0 &  \gamma\alpha^{-}_{\frac{1}{2}}  &   \frac{\Delta}{2}   &  \alpha^{+}_{-\frac{3}{2}}  \\
    0    &   0  &\alpha^{-}_{-\frac{1}{2}} &    -\frac{3\Delta}{2}
\end{bmatrix}, 
A_2= \frac{1}{2}
\begin{bmatrix}
\frac{3\Delta}{2}  &  \gamma\alpha^{+}_{\frac{1}{2}}   &   0   &    0   \\
    \gamma\alpha^{-}_{\frac{3}{2}}  &  -\frac{\Delta}{2}   &  \alpha^{+}_{-\frac{1}{2}} & 0   \\
    0 &  \alpha^{-}_{\frac{1}{2}}  &   -\frac{\Delta}{2}   &  \gamma\alpha^{+}_{-\frac{3}{2}}  \\
    0    &   0  &\gamma\alpha^{-}_{-\frac{1}{2}} &    -\frac{3\Delta}{2}
\end{bmatrix}
\end{eqnarray} 
The eigenvalues of $\mathcal{H}_{3/2}$ are
\begin{eqnarray} 
\lambda^1_{\pm}&=&\frac{1}{2}\left( \frac{\Delta}{2}+1\pm\sqrt{(\Delta-1)^{2}+3\gamma^{2}}\right),\;  
\lambda^2_\pm=\frac{1}{2}\left( \frac{\Delta}{2}-1\pm\sqrt{(\Delta+1)^{2}+3\gamma^{2}}\right),\nonumber\\ 
\lambda^3_{\pm}&=&\frac{1}{2}\left(-\frac{\Delta}{2}+\gamma \pm\sqrt{(\Delta+\gamma)^{2}+3}\right),\;
\lambda^4_\pm = \frac{1}{2}\left(-\frac{\Delta}{2}-\gamma \pm\sqrt{(\Delta-\gamma)^{2}+3}\right), 
\end{eqnarray} 
none of which are degenerate for $\Delta=0$ or $\gamma=0$. On the other hand, $\mathcal{H}_{1/2}$ also takes the form~(\ref{eq:block_diagonal_forms}) upon grouping the basis as $\{\ket{1/2}\ket{1/2}, \ket{-1/2}\ket{-1/2}\}$ and $\{\ket{1/2}\ket{-1/2}, \ket{-1/2}\ket{1/2}\}$ with the two-dimensional $A_1$ and $A_2$ matrices  given as
\begin{eqnarray}
A_1&=& \frac{1}{2}
\begin{bmatrix}
    \frac{\Delta}{2} & \gamma  \\
    \gamma  & \frac{\Delta}{2}
\end{bmatrix},
A_2= \frac{1}{2}
\begin{bmatrix}
    -\frac{\Delta}{2} & 1 \\
    1 & -\frac{\Delta}{2}
\end{bmatrix}. 
\end{eqnarray} 
The eigenvalues of $\mathcal{H}_{1/2}$ are 
\begin{eqnarray} 
\lambda^5_\pm&=&\frac{\Delta}{4} \pm \frac{\gamma}{2}, \lambda^6_\pm = -\frac{\Delta}{4}\pm \frac{1}{2},
\end{eqnarray} 
each of which is doubly degenerate due to the presence of a second identical block $\mathcal{H}_{1/2}$. 

Note that for $\Delta=0$ and for arbitrary \emph{positive} values of $\gamma$ within its allowed range, the ground state energy is given by either $\lambda^4_-$, or $\lambda^2_-$, depending on the value of $\gamma$. The eigenvectors corresponding to these two eigenvalues are given by 
\small 
\begin{eqnarray}
    \ket{\psi(\lambda^2_-)} &=& -\ket{1/2}\ket{3/2} +\frac{1+\sqrt{1+3 \gamma^2}}{\sqrt{3}\gamma}\Big(\ket{-1/2}\ket{1/2} -\ket{1/2}\ket{-1/2}\Big)+\ket{-1/2}\ket{-3/2},\nonumber\\
    \ket{\psi(\lambda_-^4)} &=&-\ket{-1/2}\ket{3/2} +\frac{\gamma+\sqrt{3+\gamma^2}}{\sqrt{3}}\Big(\ket{1/2}\ket{1/2} -\ket{-1/2}\ket{-1/2}\Big)+\ket{1/2}\ket{-3/2},
\end{eqnarray}  \normalsize 
upto normalization, both of which are eigenvectors of $\mathcal{O}$ with an eigenvalue $-1$. Similar result can be obtained for negative values of $\gamma$ as well. On the other hand, for $\gamma=0$ and $0<|\Delta|<1$, the ground state is non-degenerate with the ground state eigenvalue given by $\lambda^2_-$, and the corresponding eigenvector is 
\begin{eqnarray}
    \ket{\psi(\lambda^2_-)} &=& \frac{1}{\sqrt{2}}\Big(\ket{1/2}\ket{-1/2}-\ket{-1/2}\ket{1/2}\Big).
\end{eqnarray}

\twocolumngrid

\end{document}